\documentclass[12pt]{article}
\pdfoutput=1
\usepackage[width=420pt, height=640pt, vcentering]{geometry}
\usepackage[colorlinks=true]{hyperref}
\usepackage{amsmath}
\usepackage{amssymb}
\usepackage{amsthm}
\usepackage{graphics,graphicx}
\usepackage{lscape}
\usepackage[title,titletoc]{appendix}
\usepackage{footmisc}
\usepackage{rotating}
\usepackage{natbib}
\usepackage{enumitem}
\usepackage{textcomp}

\usepackage{bbm}

\newcommand{\p}{\rlap{\ .}}

\newcommand{\ud}{\mathrm{d}}

\newcommand{\EE}{\mathbb{E}}

\DeclareMathOperator{\Tr}{Tr}

\title{ \textbf{Unbiased Monte Carlo Simulation of Diffusion Processes}}
\author{\textbf{Louis Paulot}\thanks{Head of Quantitative Research, louis.paulot@misys.com} 
\\
[0.5cm]
{\it{Misys}\thanks{42-44 rue Washington, 75008 Paris, France}}
}
\date{May 2016}

\begin{document}
\maketitle
\begin{abstract} 
Monte Carlo simulations of diffusion processes often introduce bias in the final result, due to time discretization. Using an auxiliary Poisson process, it is possible to run simulations which are unbiased.
In this article, we propose such a Monte Carlo scheme which converges to the exact value. We manage to keep the simulation variance finite in all cases, so that the strong law of large numbers guarantees the convergence. Moreover, the simulation noise is a decreasing function of the Poisson process intensity. Our method handles multidimensional processes with nonconstant drifts and nonconstant variance-covariance matrices. It also encompasses stochastic interest rates.

\end{abstract}



\section{Introduction}
We consider the parabolic PDE\footnote{We use the Einstein summation convention: there is a sum on indices appearing twice in a formula: $\displaystyle A_\alpha B^\alpha = \sum_{\alpha = 1}^d A_\alpha B^\alpha $. We also use the Kronecker delta $\delta^{ab}$ which is 1 if $a=b$ and 0 if $a \neq b$.}
\begin{equation}
\label{eq:PDE}
\partial_t u + \mu^\alpha \partial_\alpha u + \frac{1}{2} C^{\alpha\beta} \partial_\alpha \partial_\beta u = r u
\end{equation}
with terminal condition $u_T(x) = h(x)$ in a $(1+d)$-dimensional space with time $t$ and space variables $x$ with coordinates $x_\alpha$, $\alpha = 1\ldots d$. All coefficients may depend on $t$ and $x$.
$C$ is a positive semi-definite symmetric matrix. It can be rewritten using Cholesky decomposition as $C = \sigma \sigma^\intercal$ or
$C^{\alpha\beta} =  \delta^{ab} \sigma^\alpha{}_a \sigma^\beta{}_b$.

Feynman-Kac theorem states that the solution of this PDE is given by the expected value
\begin{equation}
u(t,x) = \EE\!\left[e^{-\int_t^T r(s,X_s) \ud s} h(X_T) \mid X_t = x \right] \p
\label{eq:expected-value}
\end{equation}
under a multidimensional diffusion process
\begin{equation*}
\ud X_t = \mu \ud t + \sigma \ud W_t \p
\end{equation*}
$W_t$ is a $d$-dimensional vector of independent standard Brownian process.
Drifts $\mu$ and volatilities $\sigma$ are stochastic variables which depend on time $t$ and state variable $X$.

The expected value \eqref{eq:expected-value} can be numerically estimated using Monte Carlo simulation.
However common schemes exhibit some bias due to the discretization of time. This happens for instance with the simplest one, the Euler scheme. One usually controls this bias through smaller timestepping, which increases the computation time.

\cite{bally2014probabilistic} introduce schemes without such bias (in the case where $r=0$). This is achieved using random time steps, where the time discretization is given by jump times of an independent Poisson process. In addition, one has to multiply paths contributions by some weights which are functions of the time discretization and of the path. However this can correspond to integrating random variables which are not always integrable or do not have finite variance, which leads to poor convergence.

\cite{henry2015exact} enhance this algorithm in two different ways. First, the weight functions are obtained for all diffusion processes in term of Malliavin weights. They also manage to keep the variance of the integrated variable finite in two cases: constant volatility and correlation (constant $C^{\alpha\beta}$) or one-dimensional processes without drift ($\mu^\alpha = 0$). The variance is controlled by the intensity of the auxilliary Poisson process but it is not monotonous: it increases when the Poisson intensity becomes too small or too high. In particular, it is not possible to increase the average number of time steps beyond some value without increasing the variance.
\newline

Our first contribution is to introduce a different Monte Carlo scheme, with smaller variance. In particular the variance becomes asymptotically a decreasing function of the Poisson intensity.

As a second contribution, we show how to make the variance finite in the general case of a process with drift and nonconstant volatility, in any dimension.

Our third contribution is to handle the case where the discount rate $r$ is stochastic.
\newline

We introduce the basic principles of this scheme in section \ref{sec:poisson}. We then detail it in the one-dimensional case in section \ref{sec:one-dimension}, with numerical evidence in section \ref{sec:numerical}. We finally explain the general mutidimensional case with stochastic discount rate in section \ref{sec:multidimensional}.

\section{Unbiased scheme}
\label{sec:poisson}

\subsection{Poisson process}

Let us consider the parabolic PDE \eqref{eq:PDE}
with terminal condition
$
u_T(x) = h(x)
$.

We can rewrite this PDE as
$$
\partial_t u_t = - \mathcal{H}_t u_t
$$
where $\mathcal{H}_t$ is the elliptic differential operator
$$
\mathcal{H}_t(x) = - r(t,x) + \mu^\alpha(t,x) \partial_\alpha + \frac{1}{2} C^{\alpha\beta}(t,x) \partial_\alpha \partial_\beta \p
$$
The evolution operator can be expressed using a time ordered exponential
$$
U_{t,T} = \mathcal{P} e^{\textstyle \int_t^T \!\mathcal{H}_s \,\ud s}
$$
which is a notation for the infinite product
$$
U_{t,T} = \lim_{n \rightarrow \infty} \prod_{k=0}^{n-1} e^{\textstyle \mathcal{H}_{t+k(T-t)/n} \frac{(T-t)}{n} }
\p
$$
This allows to write the solution of equation \eqref{eq:PDE} with terminal condition $u_T(x) = h(x)$ as
$$
u_t = U_{t,T} u_T = U_{t,T} h \p
$$
Expliciting variables, this means
$$
u_t(x) = \int U_{t,T}(x,y) h(y) \, \ud y \p
$$

In some cases, for instance when the stochastic process is a pure Brownian motion or a lognormal diffusion, the resulting marginal probability measure can be computed explicitely or can be simulated exactly\footnote{
On an infinitesimal time $\delta t$, $U_{t,t+\delta t} = e^{\textstyle \mathcal{H}_t \delta t}$ acting on a test function $\phi$ can be represented by the convolution with a Gaussian kernel:
$$
\left(e^{\textstyle \mathcal{H}_t \delta t} \phi\right)\!(X_t) = e^{\textstyle -r \delta t } \frac{1}{\sqrt{(2 \pi \delta t)^d | \det(C) | }} \int \ud( \delta X)^d
\ e^{\textstyle -\frac{1}{2} \left(\frac{\delta X}{\delta t} - \mu \right)^T C^{-1} \left(\frac{\delta X}{\delta t} - \mu \right) \delta t} \phi(X_t+\delta X) \ .
$$
Defining
$$
\mathcal{L}_t = \frac{1}{2} (\dot X_t-\mu_t)^T C_t^{-1} (\dot X_t-\mu_t) + r_t
$$
with $\dot X_t = \partial_t X_t = \frac{\delta X}{\delta t}$, this convolution reads
$$
\left(e^{\textstyle \mathcal{H}_t \delta t} \phi\right)\!(X_t)  = \frac{1}{\sqrt{2 \pi | \det(C) | \delta t }} \int {\textstyle \prod\limits_i} \ud( \delta X^{(i)})
\ e^{\textstyle -\mathcal{L}_t \delta t} \phi(X_t+\delta X) \ .
$$

Integrating over infinitesimal times between $t$ and $T$, $u_t$ can thus be expressed as a Feynman path integral
$$
u_t = \mathcal{N} \int \! \mathcal{D}X_s \, e^{\textstyle  - \int_t^T \!\mathcal{L}_s \ud s} h(X_T) \p
$$
The integral is taken over all paths starting from $X_0$ with a proper normalization factor $\mathcal{N}$.
}.

In other cases, we suppose that we have a different elliptic operator
$$
\widehat{\mathcal{H}}_t = -\hat r + \hat\mu^\alpha \partial_\alpha + \frac{1}{2} \hat C^{\alpha\beta} \partial_\alpha \partial_\beta
$$
with coefficients $\hat r$, $\hat \mu$ and $\hat C$ chosen such that the PDE $\partial_t u_t = -\widehat{\mathcal{H}}_t u_t$ can be solved exactly by Monte Carlo simulation. We denote the evolution operator by $\widehat{U}_{t,t'} = \mathcal{P} e^{\textstyle \int_t^{t'} \!\widehat{\mathcal{H}}_s \,\ud s} $ and decompose $\hat C$ as  $\hat C^{\alpha\beta} = \delta^{ab} \hat\sigma^\alpha{}_a\hat\sigma^\beta{}_b$. This means that for any function $\phi(x)$ we can get the value of
$$
\left(\widehat{U}_{t,t'} \phi\right)(x) = \widehat{\EE}\!\left[e^{-\int_t^{t'} \hat{r}(s,X_s) \ud s} \phi(X_{t'}) \mid X_t = x \right]
$$
under the stochastic process
\begin{equation*}
\ud X_t = \hat\mu \ud t + \hat\sigma \ud W_t \p
\end{equation*}
With explicit coordinates this process follows the equation
\begin{equation}
\label{eq:processHat}
\ud X_t^\alpha = \hat\mu^\alpha(t,X_t) \ud t + \hat\sigma^\alpha{}_a(t,X_t) \ud W^a_t
\end{equation}
where $W^a$ are independent standard Brownian motions.

We decompose the original operator as
$$
\mathcal{H}_t = \widehat{\mathcal{H}}_t + \Delta \mathcal{H}_t
$$
with a corrective term
\begin{equation}
\label{eq:deltaH}
\Delta \mathcal{H}_t = - \Delta r + \Delta \mu^\alpha \partial_\alpha + \frac{1}{2} \Delta  C^{\alpha\beta} \partial_\alpha \partial_\beta
\end{equation}
with
\begin{eqnarray*}
\Delta r &=& r - \hat r
\\
\Delta \mu &=& \mu - \hat\mu
\\
\Delta C &=& C - \hat C \p
\end{eqnarray*}

Using this split, the evolution operator becomes
$$
U_{t,T} = \mathcal{P} e^{\textstyle \int_t^T (\widehat{\mathcal{H}}_s + \Delta \mathcal{H}_s)\,\ud s} \p
$$
On each infinitesimal time $\delta t$ we have a term
\begin{equation}
\label{eq:deltaHdt}
e^{\textstyle \Delta \mathcal{H}_t \delta t} = 1+\Delta \mathcal{H}_t \delta t = 1 - \delta t \Delta r + \delta t  \Delta\mu^\alpha \partial_\alpha + \frac{1}{2} \delta t \Delta C^{\alpha\beta}\partial_\alpha \partial_\beta \p
\end{equation}

The process with drift $\hat\mu$ and covariance matrix $\hat C$ is chosen so that it can be simulated without any bias. However the term $\Delta \mathcal{H}_t \delta t$ given in equation \eqref{eq:deltaHdt} should be computed and taken into account at any infinitesimal time, which is numerically impossible.

Instead, this contribution $\Delta \mathcal{H}_t \delta t$ is kept only with probability $\lambda_t \delta t$ over infinitesimal time $\delta t$, compensated by a factor of $\frac{1}{\lambda_t \delta t}$, such that its expected value is unchanged.
In other words, as in \cite{bally2014probabilistic} and \cite{henry2015exact} we consider a Poisson process $N_t$ with intensity $\lambda_t$. We then replace $1+\Delta \mathcal{H}_t \delta t$ by
\begin{equation}
\label{eq:jump}
1+ \delta N_t \frac{\Delta \mathcal{H}_t}{\lambda_t} \p
\end{equation}
Applied to a test function $\phi$, the expected value over the Poisson process gives the factor we want to take into account:
$$
\mathbb{E}^P\!\left[\left(1+ \delta N_t \frac{\Delta \mathcal{H}_t}{\lambda_t} \right) \phi \right] = \left(1 + \lambda_t \delta t \frac{\Delta \mathcal{H}_t}{\lambda_t}\right) \phi = e^{\textstyle \Delta \mathcal{H}_t \delta t} \phi \p
$$
In addition, the intensity $\lambda_t$ could also be a stochastic process and depend on $t$ and $X_t$.

Let $p$ be the number of Poisson jumps between times $t_0$ and $T$ and $t_k$, $k \geq 1$ the jump times. Between two Poisson jumps, the evolution operator corresponds to the diffusion process \eqref{eq:processHat}:
$$
\widehat{U}_{t_k,t_{k+1}} =  \mathcal{P} e^{\textstyle \int_{t_k}^{t_{k+1}} \!\widehat{\mathcal{H}}_s \,\ud s} \p
$$
The processes $\hat\mu_t$ and $\hat C_t$ can depend on the Poisson jump times and also on the value of $X_t$ given by the diffusion equation \eqref{eq:processHat}. Our explicit choice will be described in sections \ref{sec:one-dimension} and \ref{sec:multidimensional}.

Integrating over all times and taking the expected value on the Poisson process, we have
\begin{multline*}
u_{t_0} = \mathbb{E}^P_{t_0}\left[ \widehat{U}_{t_0,t_1} \left( 1+\frac{\Delta \mathcal{H}_{t_1}}{\lambda_{t_1}} \right)
\widehat{U}_{t_1,t_2} \left( 1+\frac{\Delta \mathcal{H}_{t_2}}{\lambda_{t_2}} \right)
\cdots
\right.
\\
\left.
\cdots
\widehat{U}_{t_{p-1},t_p} \left( 1+\frac{\Delta \mathcal{H}_{t_p}}{\lambda_{t_p}} \right)
\widehat{U}_{t_p,T} h(X_T)
\right]\ .
\end{multline*}
Operators $\widehat{U}$ act on functions by integration, which can be explicited as
\begin{multline}
\label{eq:integral}
u_{t_0}(X_{t_0}) =
\mathbb{E}^P_{t_0}\left[ \int\!\ud X_{t_1} \, \widehat{U}_{t_0,t_1}(X_t,X_{t_1})
\left( 1+\frac{\Delta \mathcal{H}_{t_1}}{\lambda_{t_1}} \right)
\right.
\\
\int\!\ud X_{t_2} \, \widehat{U}_{t_1,t_2}(X_{t_1},X_{t_2})
\left( 1+\frac{\Delta \mathcal{H}_{t_2}}{\lambda_{t_2}} \right)
\cdots
\\
\left.
\int\!\ud X_{t_p} \, \widehat{U}_{t_{p-1},t_p}(X_{t_{p-1}},X_{t_p}) 
\left( 1+\frac{\Delta \mathcal{H}_{t_p}}{\lambda_{t_p}} \right)
\int\!\ud X_T \, \widehat{U}_{t_p,T}(X_{t_p},X_T) h(X_T)
 \right] \p
\end{multline}

\subsection{Monte Carlo simulation}
In a Monte Carlo simulation, the integrals on $X_{t_1}, \cdots, X_T$ and the evolution operators $\widehat{U}$ are handled by averaging over simulated paths generated according to the law given by $\widehat{U}$. In order to get an unbiased Monte Carlo scheme, we generate random sampling for Poisson jump times $t_k$ and the values of $X_{t_k}$ at those times according to process \eqref{eq:processHat}. This process, with drift $\hat\mu$ and covariance matrix $\hat C$ is chosen so that it can be simulated exactly, \emph{i.e.} such that the Monte Carlo distribution of $X_{t_{k+1}}$ discounted at rate $\hat r$, conditional to $X_{t_k}$, tends to $\widehat{U}_{t_k,t_{k+1}}(X_{t_k},X_{t_{k+1}})$ when the number of samplings goes to infinity.

Operators $\Delta \mathcal{H}_{t_k}$ are differential operators acting on all the factors which follow in formula \eqref{eq:integral}. The first term to depend on the differentiation variable $X_{t_k}$ is in fact $\widehat{U}_{t_k,t_{k+1}}$. If we know the explicit form of this evolution kernel, we can differentiate it explicitely. This defines weights $\widehat{\mathcal{W}}_\alpha$ and $\widehat{\mathcal{W}}_{\alpha\beta}$:
\begin{eqnarray*}
\partial_{X_{t_k}^\alpha} \widehat{U}_{t_k,t_{k+1}} &=&  \widehat{\mathcal{W}}_\alpha\,  \widehat{U}_{t_k,t_{k+1}}
\\
\partial_{\partial X_{t_k}^\alpha}\partial_{X_{t_k}^\beta} \widehat{U}_{t_k,t_{k+1}} &=&  \widehat{\mathcal{W}}_{\alpha\beta} \, \widehat{U}_{t_k,t_{k+1}} \p
\end{eqnarray*}
These are similar to Malliavin weights, except that here the kernel $\widehat{U}$ includes the discount factor. With null or deterministic discount rates, $\widehat{\mathcal{W}}_\alpha$ and $\widehat{\mathcal{W}}_{\alpha\beta}$ are exactly the Malliavin weights, as introduced in this context by \cite{henry2015exact}. When the discount factor depends on $X_{t_k}$ they also incorporate the derivative of these discount factors.
However multiplying by these weights at this step would result in computing the expected value of a quantity with infinite variance. We will explain this issue and how to solve it in the following sections.

The other terms which can be functions of $X_{t_k}$ are $\hat\mu$, $\hat C$, $\hat r$ and $\lambda$, depending on the choice which is made for those functions. However the variations of these variables will not contribute to the final result.
To make this clear, we rewrite formula \eqref{eq:integral} in a recursive way for intermediate Poisson jump times:
\begin{equation}
\label{eq:recursive}
u_{t_{k-1}}(X_t) =
\mathbb{E}_{t_{k-1}}\!\left[ \int\!\ud X_{t_{k}} \, \widehat{U}_{t_{k-1},t_k}(X_{t_{k-1}},X_{t_{k}})
\left( 1+\frac{\Delta \mathcal{H}_{t_{k}}}{\lambda_{t_{k}}} \right)
u_{t_{k}}(X_{t_{k}})
 \right] \p
\end{equation}
In this formula, $\Delta\mathcal{H}_{t_k}$ acts by differentiation with respect to $X_{t_k}$ on $u_{t_{k}}(X_{t_{k}})$. The formula for $u_{t_{k}}$ involves $\hat\mu$, $\hat C$ and $\lambda$. However these functions and the process \eqref{eq:processHat} are only intermediate objects used in the computation: the final value of $u_{t_{k}}$ does not depend on the values chosen for these functions. As a consequence, their variations do not contribute to the derivatives of $u_{t_{k}}$. In the following, differentiations in operators $\Delta\mathcal{H}_{t_k}$ are therefore computed with frozen $\hat\mu$, $\hat C$ and $\lambda$. In order to make this clear, we will denote by $X_t^*$ those frozen values which should not be variated when differentiating.

\subsection{Infinite variance}

Let us consider the simple case of a pure Brownian process. The evolution operator is the Gaussian kernel
$$
U_{s,t}(W_s,W_t) = \frac{1}{\sqrt{2\pi (t-s)}} e^{\textstyle -\frac{(W_t^a-W_s^a)(W_t^a-W_s^a)}{2 (t-s)}} \p
$$
The corresponding Malliavin weights are therefore
\begin{eqnarray*}
\mathcal{W}_a &=& \frac{1}{U_{s,t}}\partial_{W_s^a} U_{s,t} = \frac{\Delta W^a}{\Delta t} \\
\mathcal{W}_{ab} &=& \frac{1}{U_{s,t}}\partial_{W_s^a}\partial_{W_s^b} U_{s,t} = \frac{\Delta W^a\Delta W^b}{\Delta t^2} -\frac{\delta_{ab}}{\Delta t}
\end{eqnarray*}
with
$\Delta W = W_t-W_s$ and $\Delta t = t-s$.

The time $\Delta t$ between two jumps is given by a Poisson law with density $\lambda e^{-\lambda \Delta t}$ which behaves as $O(1)$ at small $\Delta t$. As $\Delta W = O(\sqrt{\Delta t})$, Malliavin weights are of orders
$
\mathcal{W}_a = O\!\left( \frac{1}{\sqrt{\Delta t}}\right)
$ and $
\mathcal{W}_{ab} =O\!\left( \frac{1}{\Delta t}\right)
$. This remains true for a large class of diffusion processes. Direct multiplication by weights $\widehat{\mathcal{W}}_\alpha$ and $\widehat{\mathcal{W}}_{\alpha\beta}$ in \eqref{eq:integral} would therefore give a quantity of infinite variance. This would result in poor Monte Carlo convergence. In order for the expected value over jump time to be well defined and the integrand to have finite variance, the random variable we integrate should be of order $O\!\left(\frac{1}{\Delta t^\gamma}\right)$ with $\gamma < \frac{1}{2} $, so that its square is of order $O\!\left(\frac{1}{\Delta t^{2\gamma}}\right)$ with $2\gamma < 1 $.

In order to have a variable which behaves as $O(1)$ when $\Delta t$ is small, the weights should be multiplied by quantity of order $O(\Delta t)$. In equation \eqref{eq:integral}, $\Delta \mathcal{H}_{t_k} \widehat{U}_{t_k,t_{k+1}}(X_{t_k},\allowbreak X_{t_{k+1}})$ is multiplied by $1+\frac{\Delta \mathcal{H}_{t_{k+1}}}{\lambda}$, unless $t_k$ is the last Poisson time ($k=p$).

Let us consider first the term $\Delta \mathcal{H}_{t_{k+1}}$. According to its definition \eqref{eq:deltaH}, it is composed of terms proportional to $r(t_{k+1},X_{t_{k+1}})-\hat r(t_{k+1},X_{t_{k+1}})$, $\mu(t_{k+1},X_{t_{k+1}}) - \hat \mu(t_{k+1},\allowbreak X_{t_{k+1}})$ and $C(t_{k+1},X_{t_{k+1}})-\hat C(t_{k+1},X_{t_{k+1}})$. Our strategy is therefore to make all these terms of order $O(\Delta t_k) = O(t_{k+1} - t_k)$.

The other multiplicative term is $1$ which is not in $O(\delta t)$ and is multiplied by $\widehat{U}_{t_{k+1},t_{k+2}}(X_{t_{k+1}},X_{t_{k+2}})$. For this term, we will exploit the specific form of $\widehat{U}_{t_k,t_{k+1}}(X_{t_k},X_{t_{k+1}})$ to transfer the derivative on $X_{t_k}$ to a derivative on $X_{t_{k+1}}$. This derivative is then transferred to $\widehat{U}_{t_{k+1},t_{k+2}}(X_{t_{k+1}},X_{t_{k+2}})$ by integration by parts.

The last piece to handle is the last Poisson time, when the weights $\mathcal{W}$ must be multiplied by the final payoff. In this case, we use antithetic sampling as in \cite{henry2015exact} to have a final term in $O(\Delta t_p) = O(T-t_p)$.

Doing so, the random variable to integrate remains of order $O(1)$ and thus has finite variance under the Poisson law. Then the strong law of large numbers applies and the Monte Carlo convergence remains in $O\!\left(\frac{1}{\sqrt{N}}\right)$.

We now detail this scheme in the one-dimensional case with deterministic discount rate in section \ref{sec:one-dimension} and in the general multidimensional case with stochastic discount rate in section \ref{sec:multidimensional}.

\section{One-dimensional Monte Carlo scheme}
\label{sec:one-dimension}

Let us consider first the case of a one-dimensional process
$$
\ud S = \mu(t,S) \ud t + \sigma(t,S) \ud W \p
$$
We take a deterministic discount rate $r(t)$ and we want to price a European option of maturity $T$ with payoff $h(S_T)$. In other words, we want to solve the parabolic PDE
$$
\partial_t u_t(S) + \mu(t,S) \partial_S u_t(S) + \frac{1}{2} \sigma(t,S)^2 \partial_S^2 u_t(S) = r(t) u_t(S)
$$
with terminal condition $u_T(S) = h(S)$.

\subsection{Monte Carlo path}

We take a constant Poisson process intensity $\lambda$. Starting at date $t_0$, we draw at random the first Poisson time $t_1$. For instance, we draw a random uniform number $q$ between 0 and 1 and invert the cumulative law: $t_1 = t_0 -\frac{\log(q)}{\lambda}$. We can thus iteratively draw times $t_k$ until we get a date larger than $T$. We will denote by $p$ the last Poisson time before $T$.

Between two succesive dates $t_k$ and $t_{k+1}$ we will simulate a process with drift $\hat\mu(t,S)$ and volatility $\hat\sigma(t,S)$. More precisely, we consider a Brownian process $W_t$ and we choose a function $f_{(k)}(\Delta t, \Delta W)$ at each time $t_k$, such that $f_{(k)}(0,0) = 0$ and which can depend on $t_k$ and $S_{t_k}$. We then define 
\begin{equation}
\label{eq:Sfromf}
S_t = S_{t_k} + f_{(k)}(t-t_k, W_t-W_{t_k}) \p
\end{equation}
This corresponds to the It\^o process
$$
\ud S_t = \left( \partial_{\Delta t} f_{(k)} + \frac{1}{2} \partial_{\Delta W}^2 f_{(k)} \right) \ud t + \partial_{\Delta W} f_{(k)} \, \ud W_t \p
$$
Identifying with
$$
\ud S_t = \hat\mu \ud t + \hat\sigma \ud W_t
$$
we define
\begin{eqnarray}
\label{eq:hatMuDef}
\hat\mu &=& \partial_{\Delta t} f_{(k)} + \frac{1}{2} \partial_{\Delta W}^2 f_{(k)}
\\
\label{eq:hatSigmaDef}
\hat \sigma &=& \partial_{\Delta W} f_{(k)} \p
\end{eqnarray}

We suppose that we can write $f_{(k)}$ as a power series
$$
f_{(k)}(\Delta t, \Delta W) = \sum_{i,j} \frac{1}{i! j!}f_{ij} \Delta t^i \Delta W^j \p
$$
($f_{ij}$ are coefficients which depend on $k$ but we do not make this explicit in order to simplify the notation.)
Then for $\Delta t = t_{k+1}-t_k$ and $\Delta W = W_{t_{k+1}} - W_{t_k}$ we have
\begin{eqnarray}
\label{eq:TaylorHatMu}
\hat\mu(t_{k+1},S_{t_{k+1}}) &=& f_{10} + f_{11} \, \Delta W + \frac{1}{2} f_{02} + \frac{1}{2} f_{03} \,\Delta W + O(\Delta t)
\\
\label{eq:TaylorHatSigma}
\hat \sigma(t_{k+1},S_{t_{k+1}}) &=& f_{01}+f_{02} \,\Delta W + O(\Delta t)
\end{eqnarray}
where we use $\Delta W = O(\sqrt{\Delta t})$.

On the other hand, a Taylor expansion of functions $\mu\big(t_k + \Delta t,S_{t_k} + f_{(k)}(\Delta t, \Delta W)\big)$ and $\sigma\big(t_k + \Delta t,S_{t_k} + f_{(k)}(\Delta t, \Delta W)\big)$ gives 
\begin{eqnarray}
\label{eq:TaylorMu}
\mu(t_{k+1},S_{t_{k+1}}) &=& \mu(t_k,S_{t_k}) + \partial_S \mu(t_k,S_{t_k}) f_{01} \, \Delta W + O(\Delta t)
\\
\label{eq:TaylorSigma}
\sigma(t_{k+1},S_{t_{k+1}}) &=& \sigma(t_k,S_{t_k}) + \partial_S \sigma(t_k,S_{t_k}) f_{01} \, \Delta W + O(\Delta t) \p
\end{eqnarray}

We have $\sigma(t_{k+1},S_{t_{k+1}}) = \hat \sigma(t_{k+1},S_{t_{k+1}}) + O(\Delta t)$ and therefore $C(t_{k+1},S_{t_{k+1}})-\hat C(t_{k+1},S_{t_{k+1}}) = \sigma(t_{k+1},S_{t_{k+1}})^2 - \hat \sigma(t_{k+1},S_{t_{k+1}})^2 = O(\Delta t)$ if and only if coefficients in equations \eqref{eq:TaylorHatSigma} and \eqref{eq:TaylorSigma} are equal. This means
\begin{eqnarray}
\label{eq:f01}
f_{01} &=& \sigma(t_k,S_{t_k})
\\
\label{eq:f02}
f_{02} &=& \partial_S \sigma(t_k,S_{t_k}) f_{01} = \sigma(t_k,S_{t_k}) \partial_S \sigma(t_k,S_{t_k}) \p
\end{eqnarray}
Similarly, we have $\mu(t_{k+1},S_{t_{k+1}}) - \hat \mu(t_{k+1},S_{t_{k+1}}) = O(\Delta t)$ if and only if
\begin{eqnarray*}
f_{10} + \frac{1}{2} f_{02} &=& \mu(t_k,S_{t_k})
\\
f_{11} + \frac{1}{2} f_{03} &=& \partial_S \mu(t_k,S_{t_k}) f_{01} \p
\end{eqnarray*}
Using the expressions for $f_{01}$ and $f_{02}$ in these equations, this reads
\begin{eqnarray}
\label{eq:f10}
f_{10} &=& \mu(t_k,S_{t_k}) - \frac{1}{2} \sigma(t_k,S_{t_k}) \partial_S \sigma(t_k,S_{t_k})
\\
\label{eq:f11}
f_{11}  &=& \sigma(t_k,S_{t_k}) \partial_S \mu(t_k,S_{t_k})  - \frac{1}{2} f_{03}\p
\end{eqnarray}
In addition, the continuity of $S_t$ at $t=t_k$ in \eqref{eq:Sfromf}, equivalent to $f(0,0)=0$, gives
\begin{equation}
\label{eq:f00}
f_{00} = 0 \p
\end{equation}

Except expressions \eqref{eq:f01}, \eqref{eq:f02}, \eqref{eq:f10}, \eqref{eq:f11} and \eqref{eq:f00} we have the freedom to choose all other $f_{ij}$ coefficients. One simple choice is to set them to 0: $f_{ij} = 0$ for $(i=0, j\geq 3)$, $(i=1, j \geq 2)$ and $(i \geq 2)$. Between $t_k$ and $t_{k+1}$ we thus choose the function $f$ to be
\begin{multline}
f_{(k)}(\Delta t, \Delta W) = \mu(t_k,S_{t_k}) \Delta t + \sigma(t_k,S_{t_k}) \Delta W 
\\
+ \frac{1}{2} \sigma(t_k,S_{t_k}) \partial_S \sigma(t_k,S_{t_k}) (\Delta W^2 - \Delta t)
+  \sigma(t_k,S_{t_k}) \partial_S \mu(t_k,S_{t_k}) \Delta t \Delta W \ .
\label{eq:f1d}
\end{multline}

At time $t_k$ we draw a Gaussian variable $\Delta W_k = W_{t_{k+1}} - W_{t_k}$ with variance $\Delta t_k = t_{k+1}-t_k$ and then get recursively the Monte Carlo path
\begin{equation}
S_{t_{k+1}} = S_{t_k} + f_{(k)}(\Delta t_k, \Delta W_k) \p
\label{eq:MCpath}
\end{equation}

\subsection{Corrective terms}
\label{sec:1dcorrective}

Applying equations \eqref{eq:hatMuDef} and \eqref{eq:hatSigmaDef},
our choice \eqref{eq:f1d} corresponds to a process with drift and volatility
\begin{eqnarray*}
\hat\mu\big(t_k+\Delta t,S_{t_k}+f_{(k)}(\Delta t, \Delta W) \big)
&=&\mu(t_k,S_{t_k}) + \sigma(t_k,S_{t_k}) \partial_S \mu(t_k,S_{t_k}) \Delta W
\\
\hat\sigma\big(t_k+\Delta t,S_{t_k}+f_{(k)}(\Delta t, \Delta W)\big)
&=& \sigma(t_k,S_{t_k}) \big[ 1 +  \partial_S \sigma(t_k,S_{t_k}) \Delta W 
\\
&&\qquad\qquad\qquad\qquad\qquad
+ \partial_S \mu(t_k,S_{t_k}) \Delta t  \big] \p
\end{eqnarray*}
As the discount rate $r(t)$ is deterministic, we also take
$$
\hat r(t,S_t) = r(t) \p
$$
Using these three function for $k-1$ with $\Delta t_k = \Delta t_{k-1} = t_k - t_{k-1}$ and $\Delta W = \Delta W_{k-1} = W_k - W_{k-1}$
and according to the definition of $\Delta \mathcal{H}_t$ we have 
$$
1+\frac{\Delta\mathcal{H}_{t_k}}{\lambda} = 1+ \frac{\Delta \mu_k}{\lambda} \partial_{S_{t_k}} + \frac{1}{2}\frac{\Delta C_k}{\lambda} \partial_{S_{t_k}}^2
$$
with
\begin{eqnarray*}
\Delta \mu_k &=& \mu(t_k,S_{t_k} )
- \hat\mu\big(t_{k-1}+\Delta t_{k-1},S_{t_{k-1}}+f(\Delta t_{k-1}, \Delta W_{k-1}) \big)
\\
\Delta C_k &=& \sigma(t_k,S_{t_k})^2
- \hat\sigma\big(t_{k-1}+\Delta t_{k-1},S_{t_{k-1}}+f(\Delta t_{k-1}, \Delta W_{k-1}) \big)^2 \p
\end{eqnarray*}
With our explicit choice of functions $f_{(k)}$, we get
\begin{eqnarray}
\nonumber
\Delta \mu_k 
&=& \mu(t_k,S_{t_k}) - \big[ \mu(t_{k-1},S_{t_{k-1}}) + \sigma(t_{k-1},S_{t_k}) \partial_S \mu(t_{k-1},S_{t_{k-1}}) \Delta W_{k-1} \big]
\\
\label{eq:deltaMudeltaC}
\Delta C_k 
&=& \sigma(t_k,S_{t_k})^2 - \sigma(t_{k-1},S_{t_{k-1}})^2 \big[  1+ \partial_S \sigma(t_{k-1},S_{t_{k-1}}) \Delta W_{k-1} 
\\
&& \qquad\qquad\qquad\qquad\qquad\qquad\qquad\qquad\qquad
+ \partial_S \mu(t_{k-1},S_{t_{k-1}}) \Delta t_{k-1} \big]^2 \p
\nonumber
\end{eqnarray}

\subsubsection{Intermediate times}
\label{sec:1dintermediate}

If $t_k$ is not the last Poisson time, \emph{i.e} $k<p$, $\Delta \mathcal{H}_k$ acts on
\begin{equation}
\label{eq:intUcorrective}
 u_k(t_k,S_{t_k}) = \int \ud S_{t_{k+1}} \widehat{U}_{t_k,t_{k+1}}(S_{t_k},S_{t_{k+1}}) \left(1+\frac{\Delta \mathcal{H}_{k+1}}{\lambda}\right) u_{t_{k+1}}(S_{t_{k+1}}) \p
\end{equation}
More generally we will consider the action on this expression of a second order differential operator
$$
\mathcal{A}_k = 1 + A_k^S \partial_{S_{t_k}} + \frac{1}{2}A_k^{SS} \partial_{S_{t_k}}^2 \p
$$

In order to compute the derivatives of expression \eqref{eq:intUcorrective} with respect to $S_{t_k}$ we consider the change of variable from $(t,S)$ to $(t,W)$ defined as in equation \eqref{eq:MCpath}:
\begin{equation*}
S = S_* + f_{(k)}(t-t_k,W-W_*) \p
\end{equation*}
with $S_* = S_{t_k}$ and $W_* = W_{t_k}$. We introduce  $S_*$ and $W_*$ notations to make clear that once the function $f$ is chosen, $S_*$ and $W_*$ are constant and should not be differentiated.
This change of variable induces for the first derivative
\begin{equation}
\label{eq:changederW}
\partial_W = ( \partial_W S ) \partial_S = \partial_{\Delta W} f_{(k)}(t-t_k,W-W_{t_k}) \, \partial_S  = \hat\sigma \partial_S \p
\end{equation}
Differentiating a second time with respect to $W$ we have
\begin{equation}
\label{eq:changederWW}
\partial_W^2 = \hat\sigma^2 \partial_S^2 + (\partial_W \hat \sigma) \partial_S = \hat\sigma^2 \partial_S^2 + (\partial_{\Delta W}^2 f_{(k)}) \partial_S \p
\end{equation}
With our particular choice \eqref{eq:f1d} for the function $f$ this is
$$
\partial_W^2 = \hat\sigma^2 \partial_S^2 +  \sigma(t_k,S_{t_k}) \partial_S \sigma(t_k,S_{t_k}) \partial_S \p
$$
Inverting these equations we have
\begin{eqnarray*}
\partial_S &=& \frac{1}{\hat \sigma} \partial_W
\\
\partial_S^2 &=& \frac{1}{\hat \sigma^2} \partial_W^2 - \frac{\partial_W \hat\sigma}{\hat\sigma^3}\partial_W \p
\end{eqnarray*}

Thus we can write in term of variable $W$
$$
\mathcal{A}_k = 1 + \frac{A_k^S}{ \hat \sigma(t_k^+,W_{t_k})} \partial_W + \frac{1}{2} \frac{A_k^{SS}}{\hat\sigma(t_k^+,W_{t_k})^2}\left[ \partial_W^2 - \frac{\partial_W \hat\sigma(t_k^+,W_{t_k})}{\hat\sigma(t_k^+,W_{t_k})} \partial_W \right] \p
$$
Using again
\begin{eqnarray*}
\hat\sigma(t_k^+,W_{t_k}) &=& \sigma(t_k,S_{t_k})
\\
\partial_W \hat\sigma(t_k^+,W_{t_k}) &=& \sigma(t_k,S_{t_k}) \partial_S\sigma(t_k,S_{t_k}) \p
\end{eqnarray*}
this becomes
\begin{equation}
\label{eq:deltaHoverLambda}
\mathcal{A}_k = 1 + \left[\frac{A_k^S}{\sigma(t_k,S_{t_k})} - \frac{1}{2} \frac{A_k^{SS} \partial_S\sigma(t_k,S_{t_k})}{\sigma(t_k,S_{t_k})^2}\right] \partial_W + \frac{1}{2} \frac{A_k^{SS}}{\sigma(t_k,S_{t_k})^2} \partial_W^2 \p
\end{equation}

After the change of variables from $(t,S)$ to $(t,W)$, equation \eqref{eq:intUcorrective} becomes
\begin{equation}
\label{eq:intUcorrectiveW}
 u_k(t_k,W_{t_k}) = \int \ud W_{t_{k+1}} \widehat{U}^{(W)}_{t_k,t_{k+1}}(W_{t_k},W_{t_{k+1}}) \left(1+\frac{\Delta \mathcal{H}_{k+1}}{\lambda}\right) u_{t_{k+1}}(W_{t_{k+1}}) \p
\end{equation}
As $W$ is a Brownian motion, in term of the new variable, the evolution operator $\widehat{U}^{(W)}$ is the product of the discount factor by a Gaussian kernel:
$$
\widehat{U}^{(W)}_{t_k,t_{k+1}}(W_{t_k},W_{t_{k+1}}) = e^{\textstyle -\int_{t_k}^{t_{k+1}} r(s) \ud s} \varphi(t_{k+1}-t_k, W_{t_{k+1}}-W_{t_k})
$$
with
\begin{equation}
\label{eq:GaussKernel}
\varphi(\Delta t, \Delta W) = \frac{1}{\sqrt{2\pi \Delta t}} e^{\textstyle - \frac{\Delta W^2}{2 \Delta t}} \p
\end{equation}

We have to compute the effect of the differential operator $ \mathcal{A}_k $ acting on $u_k$ given in formula \eqref{eq:intUcorrectiveW}, which we expand as
\begin{multline}
\label{eq:intUcorrectiveWexpanded}
 u_k(t_k,W_{t_k}) = \int \ud W_{t_{k+1}} \widehat{U}^{(W)}_{t_k,t_{k+1}}(W_{t_k},W_{t_{k+1}}) u_{t_{k+1}}(W_{t_{k+1}}) 
 \\
 + \int \ud W_{t_{k+1}} \widehat{U}^{(W)}_{t_k,t_{k+1}}(W_{t_k},W_{t_{k+1}})\frac{\Delta \mathcal{H}_{k+1}}{\lambda} u_{t_{k+1}}(W_{t_{k+1}}) 
 \p
\end{multline}

When acting on the second term in this formula, which contains $\frac{\Delta \mathcal{H}_{k+1}}{\lambda}$, the derivatives on $W_{t_k}$ are replaced by Malliavin weights:
\begin{eqnarray*}
\partial_{W_{t_k}} \widehat{U}^{(W)}_{t_k,t_{k+1}}(W_{t_k},W_{t_{k+1}}) &=& \mathcal{W}_W(\Delta t_k, \Delta W_k) \widehat{U}^{(W)}_{t_k,t_{k+1}}(W_{t_k},W_{t_{k+1}})
\\
\partial_{W_{t_k}}^2 \widehat{U}^{(W)}_{t_k,t_{k+1}}(W_{t_k},W_{t_{k+1}}) &=& \mathcal{W}_{WW}(\Delta t_k, \Delta W_k) \widehat{U}^{(W)}_{t_k,t_{k+1}}(W_{t_k},W_{t_{k+1}})
\end{eqnarray*} 
where the explicit form of the Gaussian kernel \eqref{eq:GaussKernel} gives
\begin{eqnarray*}
\mathcal{W}_W(\Delta t_k, \Delta W_k) &=& \frac{\Delta W_k}{\Delta t_k}
\\
\mathcal{W}_{WW}(\Delta t_k, \Delta W_k) &=& \frac{\Delta W_k^2 - \Delta t_k}{\Delta t_k^2} \p
\end{eqnarray*} 

For the derivatives of the first term of expression \eqref{eq:intUcorrectiveWexpanded} with respect to $W_{t_k}$, we use the symmetry of the Gaussian kernel $\varphi$, therefore of $\widehat{U}^{(W)}$, with respect to $W_{t_k}$ and $W_{t_{k+1}}$ to write
$$
\partial_{W_{t_k}} \widehat{U}^{(W)}_{t_k,t_{k+1}}(W_{t_k},W_{t_{k+1}}) = - \partial_{W_{t_{k+1}}} \widehat{U}^{(W)}_{t_k,t_{k+1}}(W_{t_k},W_{t_{k+1}}) \p
$$
Then we integrate by part in first term of  \eqref{eq:intUcorrectiveWexpanded} to get
\begin{multline*}
\partial_{W_{t_k}}  \int \ud W_{t_{k+1}} \widehat{U}^{(W)}_{t_k,t_{k+1}}(W_{t_k},W_{t_{k+1}}) u_{t_{k+1}}(W_{t_{k+1}}) =
\\
\int \ud W_{t_{k+1}} \widehat{U}^{(W)}_{t_k,t_{k+1}}(W_{t_k},W_{t_{k+1}}) \partial_{W_{t_{k+1}}}\! u_{t_{k+1}}(W_{t_{k+1}})
\end{multline*}
and similarly for the second derivative.
Then we come back from $W$ to $S$ variable at time $t_{k+1}$ using \eqref{eq:changederW} and \eqref{eq:changederWW} which in our case read
\begin{eqnarray*}
\partial_{W_{t_{k+1}}} &=& \sigma(t_k,S_{t_k}) \big[ 1 +  \partial_S \sigma(t_k,S_{t_k}) \Delta W _k
+ \partial_S \mu(t_k,S_{t_k}) \Delta t_k  \big] \partial_{S_{t_{k+1}}}
\\
\partial_{W_{t_{k+1}}}^2 &=& \sigma(t_k,S_{t_k})^2 \big[ 1 +  \partial_S \sigma(t_k,S_{t_k}) \Delta W _k
+ \partial_S \mu(t_k,S_{t_k}) \Delta t_k  \big]^2 \partial_{S_{t_{k+1}}}^2
\\
&& \qquad\qquad\qquad\qquad\qquad\qquad\qquad\qquad
+  \sigma(t_k,S_{t_k})  \partial_S \sigma(t_k,S_{t_k}) \partial_{S_{t_{k+1}}} \p
\end{eqnarray*}

Combining all terms, we finally obtain
\begin{multline*}
\mathcal{A}_k u_k = \int \ud S_{t_{k+1}} \bigg( 1 + \bigg[\frac{A_k^S}{\sigma(t_k,S_{t_k})} - \frac{1}{2}\frac{A_k^{SS} \partial_S\sigma(t_k,S_{t_k})}{\sigma(t_k,S_{t_k})^2}\bigg] \frac{\Delta W_k}{\Delta t_k}
\\
\qquad\qquad\qquad
+ \frac{1}{2}\frac{A_k^{SS}}{\sigma(t_k,S_{t_k})^2} \frac{\Delta W_k^2 - \Delta t_k}{\Delta t_k^2} \bigg) \widehat{U}_{t_k,t_{k+1}}(S_{t_k},S_{t_{k+1}}) \frac{\Delta \mathcal{H}_{k+1}}{\lambda} u_{t_{k+1}}(S_{t_{k+1}})
\\
+ \int \ud S_{t_{k+1}} \widehat{U}_{t_k,t_{k+1}}(S_{t_k},S_{t_{k+1}})
\bigg( 1 + \bigg[A_k^S - \frac{1}{2}\frac{A_k^{SS} \partial_S\sigma(t_k,S_{t_k})}{\sigma(t_k,S_{t_k})}\bigg]
\qquad\qquad
\\
\big[ 1 +  \partial_S \sigma(t_k,S_{t_k}) \Delta W _k
+ \partial_S \mu(t_k,S_{t_k}) \Delta t_k  \big] \partial_{S_{t_{k+1}}} 
\\
+ \frac{1}{2}A_k^{SS} \big[ 1 +  \partial_S \sigma(t_k,S_{t_k}) \Delta W _k + \partial_S \mu(t_k,S_{t_k}) \Delta t_k  \big]^2 \partial_{S_{t_{k+1}}} ^2 
\\
+ \frac{1}{2}\frac{A_k^{SS} \partial_S\sigma(t_k,S_{t_k})}{\sigma(t_k,S_{t_k})} \partial_{S_{t_{k+1}}}
\bigg)
u_{t_{k+1}
}(S_{t_{k+1}}) \p
\end{multline*}

We can rewrite this as
$$
\mathcal{A}_k u_k = \int \ud S_{t_{k+1}} \widehat{U}_{t_k,t_{k+1}}(S_{t_k},S_{t_{k+1}}) \mathcal{A}_{k+1} u_{t_{k+1}}(S_{t_{k+1}})
$$
with
$$
\mathcal{A}_{k+1}= 1 + A_{k+1}^S \partial_{S_{t_{k+1}}} + \frac{1}{2} A_{k+1}^{SS} \partial_{S_{t_{k+1}}}^2
$$
\begin{eqnarray}
\label{eq:recursiveA}
A_{k+1}^S &=& \big[ 1 + b_k\big] A_k^S  - \frac{1}{2} b_k \frac{ \partial_S\sigma(t_k,S_{t_k})}{\sigma(t_k,S_{t_k})} A_k^{SS}
+ d_k(\Delta t_k, \Delta W_k) \frac{\Delta \mu_{k+1}}{\lambda}
\\
A_{k+1}^{SS}&=& \big[ 1 + b_k\big]^2 A_k^{SS} + d_k(\Delta t_k, \Delta W_k) \frac{\Delta C_{k+1}}{\lambda}
\nonumber
\end{eqnarray}
and
$$
b_k = \partial_S \sigma(t_k,S_{t_k}) \Delta W_k + \partial_S \mu(t_k,S_{t_k}) \Delta t_k
$$
\begin{multline}
d_k(\Delta t,\Delta W) = 1 + \bigg[\frac{A_k^S}{\sigma(t_k,S_{t_k})} - \frac{1}{2}\frac{A_k^{SS} \partial_S\sigma(t_k,S_{t_k})}{\sigma(t_k,S_{t_k})^2}\bigg] \frac{\Delta W}{\Delta t}
\\
+ \frac{1}{2}\frac{A_k^{SS}}{\sigma(t_k,S_{t_k})^2} \frac{\Delta W^2 - \Delta t}{\Delta t^2} \ .
\label{eq:dk}
\end{multline}

Using formulas \eqref{eq:recursiveA}, we can recursively accumulate corrective terms. We start with $\mathcal{A}_0 = 1$ at $t=t_0$. Then starting from each date $t_k$ we simulate $S_{t_{k+1}}$ as given by equation \eqref{eq:MCpath} and we compute $\mathcal{A}_{k+1}$ as defined above, up to the last Poisson time $t_p$.

\subsubsection{Payoff}
\label{sec:payoff}

On the final Poisson time $t_p$, $\mathcal{A}_p$ should act on
$$
u_{t_p}(S_{t_p}) = \int \ud S_T\, \widehat{U}_{t_p,T}(S_{t_p},S_T) h(S_T)
$$
where $h$ is the payoff function.

A naive approach would be to the following. One simulates $S_T$ from $S_{t_p}$ as given in equation \eqref{eq:MCpath}:
$$
S_T = S_{t_p} + f_{(p)}(\Delta t_p, \Delta W_p)
$$
with $\Delta t_p = T-t_p$ and $\Delta W_p = W_T - W_{t_p}$ a Gaussian variable of variance $\Delta t_p$. Then one computes the payoff $h(S_T)$ and computes the derivatives using Malliavin weights. This means multiplying the discounted payoff by $d_p(\Delta t_p, \Delta W_p)$ given in equation \eqref{eq:dk}:
$$
P_T = d_p(\Delta t_p, \Delta W_p) h\big(S_T\big) \p
$$
However this term behaves as $O\!\left(\frac{1}{\Delta t_p}\right)$ at small $\Delta t_p$ and not $O(1)$ as we want.

Instead, we will use antithetic sampling on this last time step, as in \cite{henry2015exact}. More precisely, we compute
\begin{eqnarray}
\nonumber
S_T^{(+)} &=& S_{t_p} + f_{(p)}(\Delta t_p, \Delta W_p)
\\
S_T^{(0)} &=& \hat{\mathbb{E}}_{t_p}\big[ S_T \big] = S_{t_p} + \mu(t_p,S_{t_p}) \Delta t_p
\label{eq:S+S-S0}
\\
S_T^{(-)} &=& S_{t_p} + f_{(p)}(\Delta t_p, -\Delta W_p)
\nonumber
\end{eqnarray}
and then
\begin{multline*}
P_T = \frac{1}{2} d_p(\Delta t_p, \Delta W_p) h\big(S_T^{(+)}\big)
+ \frac{1}{2} d_p(\Delta t_p, -\Delta W_p) h\big(S_T^{(-)}\big)
\\
+ \left[1-\frac{1}{2} d_p(\Delta t_p, \Delta W_p) - \frac{1}{2} d_p(\Delta t_p, -\Delta W_p)\right] h\big(S_T^{(0)}\big) 
\end{multline*}
which simplifies to
\begin{multline}
\label{eq:pathContrib}
P_T =  \frac{1}{2} d_p(\Delta t_p, \Delta W_p) h\big(S_T^{(+)}\big)
+ \frac{1}{2} d_p(\Delta t_p, -\Delta W_p) h\big(S_T^{(-)}\big)
\\
- \frac{1}{2} \frac{A_p^{SS}}{\sigma(t_p,S_{t_p})^2} \frac{\Delta W_p^2 - \Delta t_p}{\Delta t_p^2} h\big(S_T^{(0)}\big) \ .
\end{multline}
The last term in $h\big(S_T^{(0)}\big)$  has expected value $0$ and the two first terms are antithetic contributions to the option price.

If the payoff function $h$ is smooth and has a Taylor expansion, one can check that $P_T$ is of order $O(1)$. This is not true for a Call or Put option payoff in the vicinity of the strike. In this case we have $P_T = O\!\left(\frac{1}{\sqrt{\Delta t_p}}\right)$. However, the probability to have the strike between $S_T^{(-)}$ and $S_T^{(+)}$ at the last time step scales as $O(\sqrt{\Delta t_p})$. As a consequence, the variance remains finite.

We finally discount on all time steps. As we take a deterministic discount rate in this first case, this factors out as a multiplication by
$
e^{\textstyle - \int_{t_0}^T r(t) \ud t} \p
$

\subsection{Monte Carlo scheme summary}

We now sum up the whole Monte Carlo scheme in this one-dimensional case with deterministic discount factor. We consider a European option with maturity $T$ and payoff $h(S_T)$.
\newline

For each path, we do the following:
\begin{enumerate}
\item Start from $S = S_0$ at time $t=t_0$. Define
\begin{eqnarray*}
A_0^S &=& 0
\\
A_0^{SS} &=& 0 \p
\end{eqnarray*}
\item On each date $t_k$
\begin{enumerate}
\item
Draw the next Poisson time $t_{k+1} = t_k + \Delta t_k$ with intensity $\lambda$. For example draw a random uniform variable $q$ between 0 and 1 and take $\Delta t_k = -\frac{\log(q)}{\lambda}$.
\\
If $t_{k+1} > T$, set $p=k$ and go to step 3.
\item
Draw a Gaussian variable $\Delta W_k$ with variance $\Delta t_k$. Get $S_{t_{k+1}}$ by equations \eqref{eq:f1d} and \eqref{eq:MCpath}:
$$
S_{t_{k+1}} = S_{t_k} + f_{(k)}(\Delta t_k, \Delta W_k) \p
$$
with
\begin{multline*}
f_{(k)}(\Delta t, \Delta W) = \mu(t_k,S_{t_k}) \Delta t + \sigma(t_k,S_{t_k}) \Delta W 
\\
+ \frac{1}{2} \sigma(t_k,S_{t_k}) \partial_S \sigma(t_k,S_{t_k}) (\Delta W^2 - \Delta t)
\\
+  \sigma(t_k,S_{t_k}) \partial_S \mu(t_k,S_{t_k}) \Delta t \Delta W \ .
\label{eq:f1d}
\end{multline*}
\item Compute $\Delta \mu_k$ and $\Delta C_k$ according to equation \eqref{eq:deltaMudeltaC}:
\begin{eqnarray*}
\Delta \mu_k 
&=& \mu(t_k,S_{t_k}) 
\\&& \qquad
- \big[ \mu(t_{k-1},S_{t_{k-1}}) + \sigma(t_{k-1},S_{t_k}) \partial_S \mu(t_{k-1},S_{t_{k-1}}) \Delta W_{k-1} \big]
\\
\Delta C_k 
&=& \sigma(t_k,S_{t_k})^2
- \sigma(t_{k-1},S_{t_{k-1}})^2 \big[  1+ \partial_S \sigma(t_{k-1},S_{t_{k-1}}) \Delta W_{k-1} 
\\
&& \qquad\qquad\qquad\qquad\qquad\qquad\qquad
+ \partial_S \mu(t_{k-1},S_{t_{k-1}}) \Delta t_{k-1} \big]^2 \p
\end{eqnarray*}
\item
Compute $A_{k+1}^S$ and $A_{k+1}^{SS}$ from $A_k^S$ and $A_k^{SS}$ as in equation \eqref{eq:recursiveA}:
\begin{eqnarray*}
A_{k+1}^S &=& \big[ 1 + b_k\big] A_k^S  - b_k \frac{ \partial_S\sigma(t_k,S_{t_k})}{\sigma(t_k,S_{t_k})} A_k^{SS}
+ d_k(\Delta t_k, \Delta W_k) \frac{\Delta \mu_{k+1}}{\lambda}
\\
A_{k+1}^{SS}&=& \big[ 1 + b_k\big]^2 A_k^{SS} + \frac{1}{2} d_k(\Delta t_k, \Delta W_k) \frac{\Delta C_{k+1}}{\lambda}
\end{eqnarray*}
with
$$
b_k = \partial_S \sigma(t_k,S_{t_k}) \Delta W_k + \partial_S \mu(t_k,S_{t_k}) \Delta t_k
$$
\begin{multline*}
d_k(\Delta t,\Delta W) = 1 + \bigg[\frac{A_k^S}{\sigma(t_k,S_{t_k})} - \frac{A_k^{SS} \partial_S\sigma(t_k,S_{t_k})}{\sigma(t_k,S_{t_k})^2}\bigg] \frac{\Delta W}{\Delta t}
\\
+ \frac{A_k^{SS}}{\sigma(t_k,S_{t_k})^2} \frac{\Delta W^2 - \Delta t}{\Delta t^2} \ .
\end{multline*}
\end{enumerate}
\item On time $t_p$
\begin{enumerate}
\item
Draw a Gaussian variable $\Delta W_p$ with variance $\Delta t_p = T-t_p$.
Compute $S_T^{(+)}$, $S_T^{(0)}$ and $S_T^{(-)}$ as in equation \eqref{eq:S+S-S0}:
\begin{eqnarray*}
S_T^{(+)} &=& S_{t_p} + f_{(p)}(\Delta t_p, \Delta W_p)
\\
S_T^{(0)} &=& S_{t_p} + \mu(t_p,S_{t_p}) \Delta t_p
\\
S_T^{(-)} &=& S_{t_p} + f_{(p)}(\Delta t_p, -\Delta W_p) \p
\end{eqnarray*}
\item
Get the undiscounted path contribution \eqref{eq:pathContrib} for the payoff $h(S_T)$:
\begin{multline*}
P_T =  \frac{1}{2} d_p(\Delta t_p, \Delta W_p) h\big(S_T^{(+)}\big)
+ \frac{1}{2} d_p(\Delta t_p, -\Delta W_p) h\big(S_T^{(-)}\big)
\\
- \frac{A_p^{SS}}{\sigma(t_p,S_{t_p})^2} \frac{\Delta W_p^2 - \Delta t_p}{\Delta t_p^2} h\big(S_T^{(0)}\big) \ .
\end{multline*}
\end{enumerate}
\item Multiply by the discount factor:
$$
e^{\textstyle - \int_{t_0}^T r(t) \ud t} P_T \p
$$
\end{enumerate}
We finally average over all path to get the unbiased Monte Carlo estimation of the option price.

\section{Numerical results}
\label{sec:numerical}

\subsection{Convergence}
\label{sec:convergence}

In order to test numerically our Monte Carlo scheme, we take a model for which we can have a closed formula as a reference value. We therefore choose the Black-Scholes model
$$
\ud S = \mu_0 S \ud t + \sigma_0 S \ud W_t
$$
which corresponds to local drift an volatility
\begin{eqnarray*}
\mu(t,S) &=& \mu_0 S
\\
\sigma(t,S) &=& \sigma_0 S \p
\end{eqnarray*}
There already exist an unbiased Monte Carlo scheme for this model, using $\log(S)$ as a variable. For the purpose of ours tests we ignore this and we naively apply our scheme and compare it to a Euler scheme.

We consider an underlying with spot $S_0 = 100$ and volatility $\sigma_0 = 50\%$. We suppose the interest rate and the drift are $r = \mu_0 = 5\%$. We price a Put option of maturity $T=1$~year at strike $K=80$. The Black-Scholes formula gives an option price of 7.8909.

\begin{figure}[!htbp]
\begin{center}		
\includegraphics[trim = 20mm 70mm 20mm 70mm, clip, width=.75\textwidth]{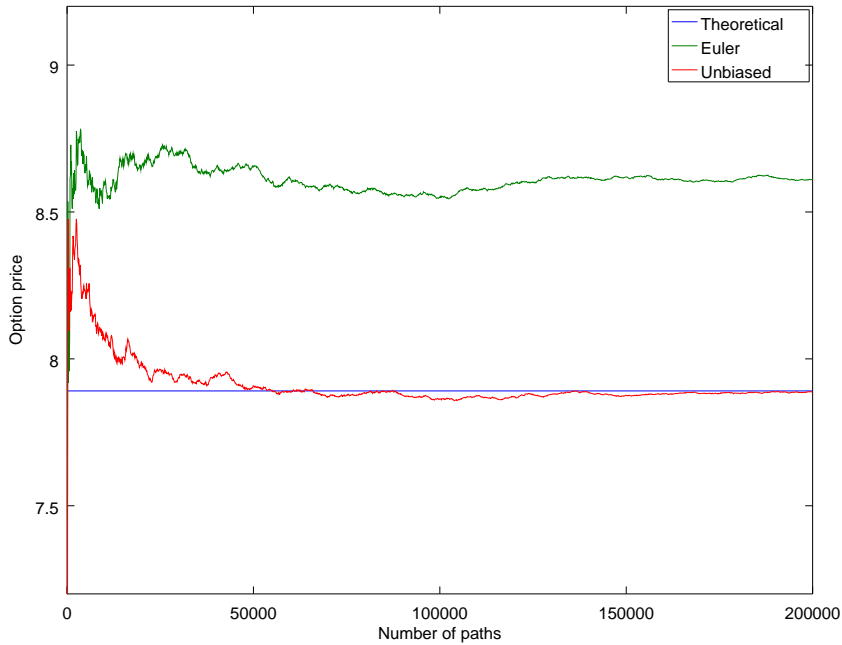}
\caption{Convergence of the simulated option price with respect to the number of paths for the unbiased scheme and the Euler scheme.}
\label{fig:convergence}
\end{center}
\end{figure}

Figure \ref{fig:convergence} shows the simulated value as a function of the number of paths for the unbiased scheme with $\lambda=3$.

For comparison, it also shows the same convergence graph for a Euler scheme with the same average number of time steps $n=4$. We see that this latter scheme converges to a biased value.

\subsection{Comparison with Euler and Milstein scheme}

When using the Euler scheme, the bias can be decreased using more time steps. This in turn increases the computation time. On the other hand, using the unbiased scheme is done at the cost of increasing the variance for small values of $\lambda$. Increasing the value of $\lambda$ makes the variance smaller but also increases the computation time, as the average number of time steps is higher.

In order to assess the performance gain which can be achieved, we consider the 80\% Put option described in the section \ref{sec:convergence} and we price it with both schemes. As the path generation has similarities with the Milstein scheme, we also compare with it. This allows to better isolate the effect of corrective terms in the unbiased scheme.
 
\begin{figure}[!htbp]
\begin{center}		
\includegraphics[width=\textwidth]{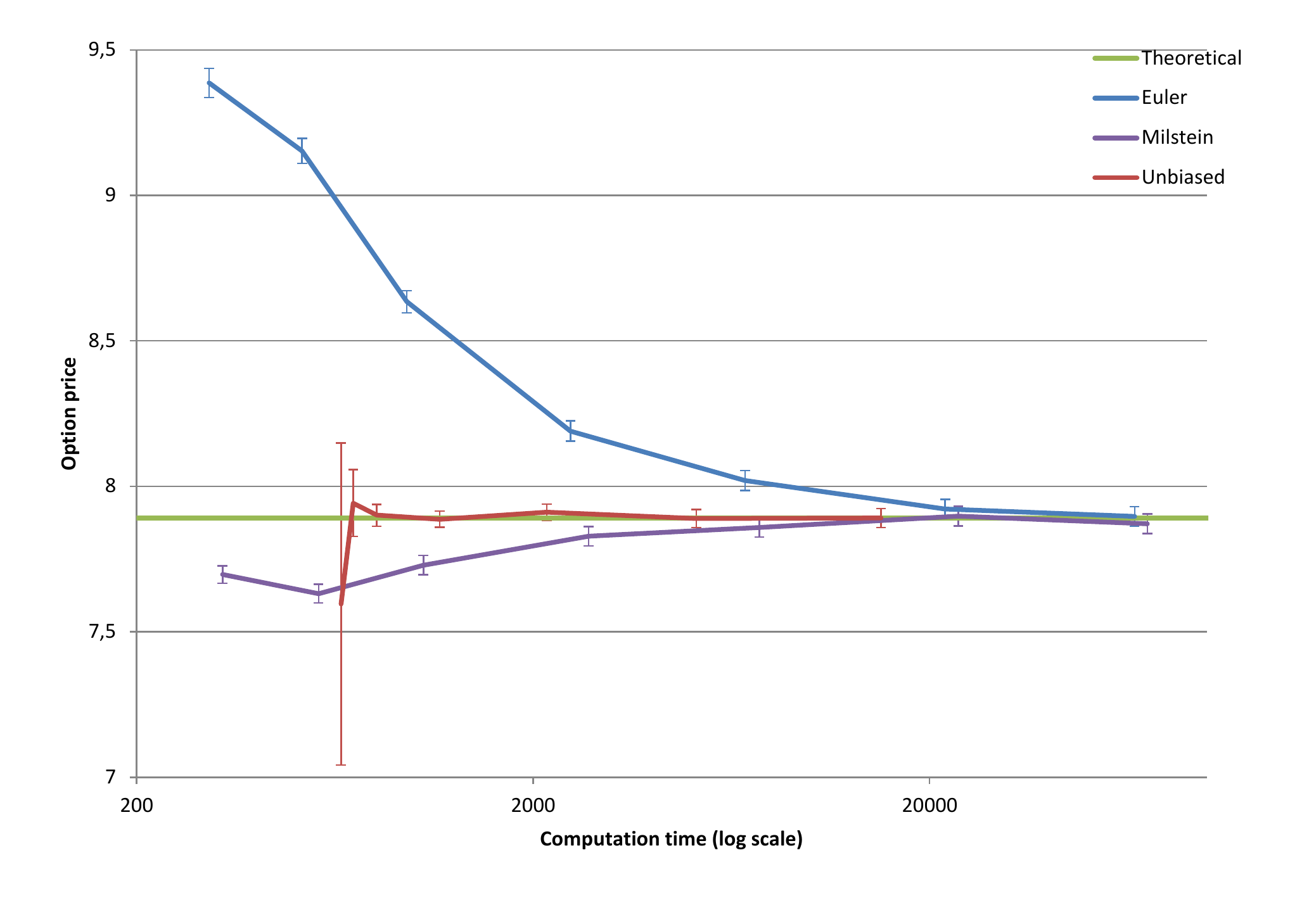}
\caption{Simulated price of a European Put obtained using the unbiased scheme, compared to Euler and Milstein schemes with the same number of Monte Carlo paths. Prices are plotted with 99\% confidence interval against the computation time in logarithimic scale.}
\label{fig:unbiased-Euler}
\end{center}
\end{figure}

For the unbiased scheme, we use increasing values of $\lambda$ from .01 to 29. For Euler and Milstein scheme, we take increasing numbers of time steps from 1 to 300. In all cases we draw $N=1$~million paths and compute numerically the estimated option price, the empirical standard deviation and the computation time.
The results are shown in figure \ref{fig:unbiased-Euler}.
The estimated prices are plotted with $99\%$ confidence interval against the computation time in logarithmic scale.

At very low Poisson intensity ($\lambda = 0.01$), the Monte Carlo variance of the unbiased scheme is higher than the Euler and Milstein schemes. However it quickly decreases when $\lambda$ increases. For $\lambda \geq 0.3$ it becomes even lower than noise of the Euler and Milstein schemes. Beyond $\lambda =1$ the Monte Carlo noise is almost stationary: the corrective terms add negligible variance compared to the basic variance of the payoff. For all values of $\lambda$, we check that the final estimate is consistent with the theoretical value, up to the Monte Carlo statistical error. The value $\lambda=1$ which appears to be optimal with respect to computation time corresponds to an average number of time steps $n=2$ over the period of 1 year.

On the opposite, the Euler scheme exhibits a large bias when the number of time steps is small. This bias decreases linearly with $\Delta T = T/n$. Performing a weighted least squares regression, we estimate the bias to behave asymptotically as $\frac{2.99}{n}$. In order to have a bias equal to the Euler Monte Carlo standard deviation $0.13$, we thus need $n \sim 230$ time steps. This corresponds to a computation time 43 times longer than the unbiased scheme with $\lambda = 1$.

The Milstein scheme has a smaller bias, asymptotically $\frac{-0.648}{n}$ in this example. We thus need $n=50$ time steps to get a bias of the same magnitude as the Monte Carlo noise. This means a computation 10 times slower than the unbiased scheme.




\section{Multidimensional process}
\label{sec:multidimensional}

We consider now the general multi-dimensional case, where all parameters can depend on $t$ and $X$ in the parabolic PDE \eqref{eq:PDE}:
\begin{equation}
\partial_t u_t(X_t) + \mu^\alpha(t,X_t) \partial_\alpha u_t(X_t) + \frac{1}{2} C^{\alpha\beta}(t,X_t) \partial_\alpha \partial_\beta u_t(X_t) = r(t,X_t) u_t(X_t) \p
\label{eq:PDEmulti}
\end{equation}
According to Feynman-Kac theorem, it corresponds to the multi-dimensional process
$$
\ud X_t^\alpha = \mu^\alpha(t,X_t) \ud t + \sigma^\alpha{}_a(t,X_t)  \ud W_t^a
$$
with stochastic discount rate $r(t,X_t)$. $\sigma(t,X_t) $ is a volatility matrix such that
$$
C^{\alpha\beta}(t,X_t)  = \sigma^\alpha{}_a(t,X_t)  \sigma^\beta{}_a(t,X_t)
$$
which can be obtained by Cholesky decomposition.
$W_t^a$ are independent standard Brownian processes.


\subsection{Monte Carlo path}

We draw Poisson times $t_k$ with intensity $\lambda$. Between these times, we simulate a $d$-dimensional Brownian process $W_t$.
This corresponds to the parabolic PDE
\begin{equation}
\label{eq:PDEBRownian}
\partial_t v_t(W_t) + \frac{1}{2} \delta^{ab} \partial_a \partial_b v_t(W_t) = 0 \p
\end{equation}

Between two Poisson times $t_k$ and $t_{k_+1}$, we will consider two change of variables\footnote{If we see the total space made of space variables and prices as a fiber bundle, this corresponds to a change of variables on the base space and the fiber.}:
\begin{itemize}
\item \textbf{Space variables.} We go from $(t,W)$ to $(t,X)$, using functions $f^\alpha_{(k)}(\Delta t, \Delta W)$:
$$
X_t^\alpha = X_{t_k}^\alpha + f^\alpha_{(k)}(t-t_k,W_t-W_{t_k}) \p
$$
\item  \textbf{Num\'eraire.} We change the num\'eraire, using a function $g_{(k)}(\Delta t, \Delta W)$
$$
u_t = v_t e^{\textstyle g_{(k)}(t-t_k,W_t-W_{t_k})} \p
$$
\end{itemize}
The change of num\'eraire transfers to derivatives as\footnote{We drop $(k)$ indices when they are not needed in order to simplify the notation.}
\begin{eqnarray*}
\partial_t v_t &=& e^{-g} \partial_t u_t - (\partial_t g) \, e^{-g} u_t
\\
\partial_a v_t &=& e^{-g} \partial_a u_t - (\partial_a g) \, e^{-g} u_t
\\
\partial_a \partial_b v_t &=& e^{-g} \partial_a \partial_b u_t - (\partial_a\partial_b g) \, e^{-g} u_t
 + (\partial_a g)(\partial_b g) \, e^{-g} u_t 
 \\ && \qquad\qquad\qquad\qquad\qquad
 - (\partial_a g) \, e^{-g} \partial_b u_t
 - (\partial_b g) \, e^{-g} \partial_a u_t \p
\end{eqnarray*}
Using the notations
$$
\partial_a = \partial_{W^a} \qquad \partial_\alpha = \partial_{X^\alpha}
$$
we define
$$
e^\alpha{}_a = \partial_a f^\alpha
$$
and
$$
c^\alpha{}_{ab} =  \partial_a \partial_b f^\alpha \p
$$
We will also use the inverse matrix $e^a{}_\alpha$, with defining properties
$$
e^\alpha{}_a e^a{}_\beta = \delta^\alpha_\beta \qquad  e^a{}_\alpha e^\alpha{}_b = \delta^a_b \p
$$
The change of space variables induces the following transformations on derivatives:
\begin{eqnarray*}
\partial_a &=& e^\alpha{}_a \, \partial_\alpha
\\
\partial_a \partial_b &=& e^\alpha{}_a e^\beta{}_b \, \partial_\alpha \partial_\beta + c^\alpha{}_{ab} \, \partial_\alpha
\\
\partial_t\!\!\mid_W &=& \partial_t\!\!\mid_X + \partial_t f^\alpha \, \partial_\alpha \p
\end{eqnarray*}

In term of the new variables $t$, $X$ and $u_t$, PDE \eqref{eq:PDEBRownian} thus becomes
\begin{equation}
\partial_t u_t + \hat\mu^\alpha \partial_\alpha u_t + \frac{1}{2} \hat C^{\alpha\beta} \partial_\alpha \partial_\beta u_t = \hat r u_t
\label{eq:PDEhat}
\end{equation}
with
\begin{eqnarray}
\hat\mu^\alpha &=& \partial_t f^\alpha - (\partial_a g) e^\alpha{}_a + \frac{1}{2} c^\alpha{}_{aa} 
\nonumber
\\
\hat C^{\alpha\beta} &=& e^\alpha{}_a e^\beta{}_a
\label{eq:hatMuCr}
\\
\hat r &=& \partial_t g  + \frac{1}{2}\partial_a\partial_a g - \frac{1}{2} (\partial_a g) (\partial_a g) \p
\nonumber
\end{eqnarray}

We want to find functions $f^\alpha_{(k)}(\Delta t, \Delta W)$ and $g_{(k)}(\Delta t, \Delta W)$ such that $\mu - \hat\mu$, $C - \hat C$ and $r - \hat r$ behaves as $O(\Delta t)$ at small $\Delta t$.

We suppose that we can expand $f^\alpha_{(k)}$ and $g_{(k)}$ as power series in $\Delta t$ and $\Delta W$:
\begin{eqnarray*}
f^\alpha_{(k)}(\Delta t, \Delta W) &=& f^\alpha_{00} + (f^\alpha_{01})_a \Delta W^a + f^\alpha_{10} \Delta t + \frac{1}{2!} (f^\alpha_{02})_{ab} \Delta W^a \Delta W^b
\\ && \qquad\qquad
+ (f^\alpha_{11})_a \Delta t\Delta W^a + \frac{1}{3!} (f^\alpha_{03})_{abc} \Delta W^a \Delta W^b \Delta W^c + \cdots
\\
g_{(k)}(\Delta t, \Delta W) &=& g_{00} + (g_{01})_a \Delta W^a + g_{10} \Delta t + \frac{1}{2!} (g_{02})_{ab} \Delta W^a \Delta W^b
\\ && \qquad\qquad
+ (g_{11})_a \Delta t\Delta W^a + \frac{1}{3!} (g_{03})_{abc} \Delta W^a \Delta W^b \Delta W^c + \cdots
\end{eqnarray*}
In this expression, $(f^\alpha_{ij})_{abc\cdots}$ or $(g_{ij})_{abc\cdots}$ are tensors which are symmetric in space indices $a,b,c, \cdots$.

The continuity contraint at $t=t_k$ is $f^\alpha(0,0) = 0$. It translates to
$$
f^\alpha_{00} = 0 \p
$$
In addition, we choose num\'eraires such that $u_t$ and $v_t$ coincide at the beginning of the period, $t=t_k$. Mathematically this is
$g(0,\Delta W) = 0$, which gives for all $j$
$$
(g_{0j}) = 0 \p
$$
Taking into account these constraints, equations \eqref{eq:hatMuCr} have Taylor expansions
\begin{eqnarray}
\hat \mu^\alpha &=& f^\alpha_{10} + (f^\alpha_{11})_a \Delta W^a + \frac{1}{2} (f^\alpha_{02})_{bb} + \frac{1}{2} (f^\alpha_{03})_{bba} \Delta W^a + O(\Delta t)
\label{eq:order1muhat}
\\
\hat C^{\alpha\beta} &=& (f^\alpha_{01})_b (f^\beta_{01})_b + (f^\alpha_{01})_b (f^\beta_{02})_{ba} \Delta W^a
+ (f^\beta_{01})_b (f^\alpha_{02})_{ba} \Delta W^a + O(\Delta t)
\label{eq:order1Chat}
\\
\hat r &=& g_{10} + (g_{11})_a \Delta W^a + O(\Delta t) \p
\label{eq:order1rhat}
\end{eqnarray}
On the other hand, a Taylor expansion of the parameters of PDE \eqref {eq:PDEmulti} gives
\begin{eqnarray}
\mu^\alpha &=& \mu^\alpha(t_k,X_{t_k}) + e^\gamma{}_a(t_k,W_{t_k}) \partial_\gamma \mu^\alpha(t_k,X_{t_k}) \Delta W^a + O(\Delta t)
\label{eq:order1mu}
\\
C^{\alpha\beta} &=& C^{\alpha\beta}(t_k,X_{t_k}) + e^\gamma{}_a(t_k,W_{t_k}) \partial_\gamma C^{\alpha\beta}(t_k,X_{t_k}) \Delta W^a + O(\Delta t)
\label{eq:order1C}
\\
r &=& r(t_k,X_{t_k}) + e^\gamma{}_a(t_k,W_{t_k}) \partial_\gamma r(t_k,X_{t_k}) \Delta W^a + O(\Delta t) \p
\label{eq:order1r}
\end{eqnarray}
From its definition, $e^\gamma{}_a$ is
$$
e^\gamma{}_a = \partial_a f^\alpha = (f^\gamma_{01})_a + (f^\gamma_{02})_{ab} \Delta W^b + O(\Delta t) \p
$$
At the beginning of the period, with $\Delta t = 0$ and $\Delta W = 0$, this is
$$
e^\gamma{}_a(t_k,W_{t_k}) = (f^\gamma_{01})_a \p
$$

We want to make $\mu-\hat\mu$, $C-\hat C$ and $r-\hat r$ vanish up to $O(\Delta t)$ terms.

Let us start with the constant term in equations \eqref{eq:order1Chat} and \eqref{eq:order1C}. From a Cholesky decomposition
$$
C^{\alpha\beta}(t_k,X_{t_k}) = \sigma^\alpha{}_b(t_k,X_{t_k}) \sigma^\beta{}_b(t_k,X_{t_k})
$$
we get a solution
\begin{equation}
\label{eq:fa01}
(f^\alpha_{01})_b = \sigma^\alpha{}_b(t_k,X_{t_k}) \p
\end{equation}
This also gives
$$
e^\alpha{}_a(t_k,W_{t_k}) = \sigma^\alpha{}_a(t_k,X_{t_k})  \p
$$
Its inverse is
$$
e^a{}_\alpha(t_k,W_{t_k}) = \big[\sigma(t_k,W_{t_k})^{-1}\big]^a{}_\alpha \p
$$
Equating the first order terms of equations \eqref{eq:order1Chat} and \eqref{eq:order1C}, we have
\begin{equation*}
e^\alpha{}_b(t_k,W_{t_k}) (f^\beta_{02})_{ba} 
+ e^\beta{}_b(t_k,W_{t_k}) (f^\alpha_{02})_{ba} = e^\gamma{}_a(t_k,W_{t_k}) \partial_\gamma C^{\alpha\beta}(t_k,X_{t_k})
\end{equation*}
where $C^{\alpha\beta}$ is symmetric in indices  $\alpha$ and $\beta$ and $(f^\beta_{02})_{ab}$ in indices $a$ and $b$.
We multiply this equation by $e^c{}_\alpha(t_k,W_{t_k})$ and $e^d{}_\beta(t_k,W_{t_k})$, also using $e^a{}_\alpha e^\alpha{}_b = \delta^a_b$:
\begin{equation}
f^d{}_{ca} + f^c{}_{da} = C_a{}^{cd}
\label{eq:efefedC}
\end{equation}
where we introduced the notations
\begin{eqnarray*}
f^a{}_{bc} &=&  e^a{}_\alpha(t_k,W_{t_k}) (f^\alpha_{02})_{bc}
\\
C_a{}^{bc} &=& e^\alpha{}_a(t_k,W_{t_k}) e^b{}_\beta(t_k,W_{t_k})e^c{}_\gamma(t_k,W_{t_k})  \partial_\alpha C^{\beta\gamma}(t_k,X_{t_k}) \p
\end{eqnarray*}
$f^a{}_{bc}$ and $C_a{}^{bc}$ are tensors symmetric in the two last indices.
We write equation \eqref{eq:efefedC} for the three cyclic permutations of the indices and get
\begin{eqnarray}
f^b{}_{ca} + f^c{}_{ba} &=& C_a{}^{bc}
\label{eq:ffC1}
\\
f^c{}_{ab} + f^a{}_{cb} &=& C_b{}^{ca}
\label{eq:ffC2}
\\
f^a{}_{bc} + f^b{}_{ac} &=& C_c{}^{ab} \p
\label{eq:ffC3}
\end{eqnarray}
The linear combination \eqref{eq:ffC3} + \eqref{eq:ffC2} - \eqref{eq:ffC1} then gives
$$
f^a{}_{bc} = \frac{1}{2} ( C_c{}^{ab} + C_b{}^{ca} - C_a{}^{bc} ) \p
$$
Inverting the definition of $f^a{}_{bc}$ in term of $(f^\alpha_{02})_{bc}$ we get
\begin{eqnarray}
(f^\alpha_{02})_{bc} &=& \frac{1}{2} \sigma^\alpha{}_a(t_k,X_{t_k})  ( C_c{}^{ab} + C_b{}^{ca} - C_a{}^{bc} ) 
\nonumber
\\
&=& \frac{1}{2} \Big[ \phantom{+}\ \sigma^\gamma{}_c(t_k,X_{t_k}) e^b{}_\beta(t_k,X_{t_k}) \partial_\gamma C^{\alpha\beta}(t_k,X_{t_k})  
\nonumber
\\
\label{eq:fa02}
&& \phantom{\frac{1}{2} \Big[}
+ \sigma^\beta{}_b(t_k,X_{t_k}) e^c{}_\gamma(t_k,X_{t_k}) \partial_\beta C^{\alpha\gamma}(t_k,X_{t_k})  
\\
\nonumber
&& \phantom{\frac{1}{2} \big[}
-  e^b{}_\beta(t_k,X_{t_k}) e^c{}_\gamma(t_k,X_{t_k}) C^{\alpha\delta}(t_k,X_{t_k}) \partial_\delta C^{\beta\gamma}(t_k,X_{t_k})  \Big]
\p
\end{eqnarray}

Equating terms in equations \eqref{eq:order1muhat} and \eqref{eq:order1mu} we get
\begin{eqnarray*}
f^\alpha_{10} &=& \mu^\alpha(t_k,W_{t_k}) - \frac{1}{2}(f^\alpha_{02})_{bb}
\\
(f^\alpha_{11})_a &=&  \sigma^\gamma{}_a(t_k,X_{t_k}) \partial_\gamma \mu^\alpha(t_k,W_{t_k}) - \frac{1}{2}(f^\alpha_{03})_{bba} \p
\end{eqnarray*}
Using the property $\Tr ( M^{-1} \partial_\alpha M) = \partial_\alpha\! \log(\det(M))$ we rewrite $(f^\alpha_{02})_{bb}$ as
$$
(f^\alpha_{02})_{bb} = \partial_\beta C^{\alpha\beta}(t_k,W_{t_k}) - \frac{1}{2} C^{\alpha\gamma}(t_k,W_{t_k}) \partial_\gamma \!\log(\det(C))(t_k,W_{t_k})
$$
and get
\begin{equation}
\label{eq:fa10}
f^\alpha_{10} = \mu^\alpha(t_k,W_{t_k}) - \frac{1}{2} \partial_\beta C^{\alpha\beta}(t_k,W_{t_k}) 
+\frac{1}{4} C^{\alpha\gamma}(t_k,W_{t_k}) \partial_\gamma \!\log(\det(C))(t_k,W_{t_k}) \ .
\end{equation}
Making the additional choice $f^\alpha_{ij} = 0$ for $f_{ij} = 0$ for $(i=0, j\geq 3)$, $(i=1, j \geq 2)$ and $(i \geq 2)$, the expression for $(f^\alpha_{11})_a$ simplifies to
\begin{equation}
\label{eq:fa11}
(f^\alpha_{11})_a =  \sigma^\gamma{}_a(t_k,X_{t_k}) \partial_\gamma \mu^\alpha(t_k,W_{t_k}) \p
\end{equation}
This completes the definition of $f_{(k)}^\alpha$ as a polynomial in $\Delta t$ and $\Delta W$
$$
f^\alpha_{(k)}(\Delta t, \Delta W) = (f^\alpha_{01})_a \Delta W^a + f^\alpha_{10} \Delta t + \frac{1}{2} (f^\alpha_{02})_{ab} \Delta W^a \Delta W^b 
+ (f^\alpha_{11})_a \Delta t\Delta W^a
$$
with all coefficients defined in equations \eqref{eq:fa01}, \eqref{eq:fa10}, \eqref{eq:fa02} and \eqref{eq:fa11}.

Finally, equating equations \eqref{eq:order1rhat} and \eqref{eq:order1r} we get
\begin{eqnarray}
\label{eq:g10}
g_{10} &=& r(t_k,X_{t_k})
\\
\label{eq:g11}
(g_{11})_a &=& \sigma^\gamma{}_a(t_k,X_{t_k})  \partial_\gamma r(t_k,X_{t_k}) \p
\end{eqnarray}
Choosing all other $g_{ij}$ coefficients to be 0, we also get $g_{(k)}$ as a polynomial in $\Delta t$ and $\Delta W$
$$
g_{(k)}(\Delta t, \Delta W) = g_{10} \Delta t +  (g_{11})_a \Delta t \Delta W^a \p
$$

Using function $f^\alpha_{(k)}$ and $g_{(k)}$ we are able to solve the PDE \eqref{eq:PDEhat} by Monte Carlo simulation. At time $t_k$ we draw $d$ independent Gaussian variable $\Delta W^a_k, 1 \geq a \geq d$ with variance $\Delta t_k = t_{k+1}-t_k$. Then we compute the value of the system at the following date
$$
X^\alpha_{t_{k+1}} = X^\alpha_{t_k} + f^\alpha_{(k)}(\Delta t_k,\Delta W_k) \p
$$
The (stochastic) discount factor between $t_k$ and $t_{k+1}$ is
$$
D(t_k,t_{k+1}) = e^{\textstyle -g_{(k)}(\Delta t_k,\Delta W_k)} \p
$$
We reccursively multply it to get the discount factor up from $t_0$ to $t_{k+1}$ as
$$
D_{k+1} =D_k D(t_k,t_{k+1}) = D_k  e^{\textstyle -g_{(k)}(\Delta t_k,\Delta W_k)} \rlap{\ ,}
$$
starting from $D_0 = 1$.

\subsection{Corrective terms}

From the definition of $f^\alpha_{(k)}$ and $g_{(k)}$, between $t_k$ ad $t_{k+1}$, 
we have
$$
e^\alpha_{(k)a}(\Delta t, \Delta W) = (f^\alpha_{01})_a + (f^\alpha_{02})_{ab} \Delta W^b + (f^\alpha_{11})_a \Delta t
$$
Using the definitions of coefficients, this is
\begin{equation}
\label{eq:edtdw}
e^\alpha_{(k)a}(\Delta t, \Delta W) = \sigma^\alpha{}_a(t_k,X_{t_k}) 
+  c^\alpha_{(k)ab} \Delta W^b
+\sigma^\gamma{}_a(t_k,X_{t_k}) \partial_\gamma \mu^\alpha(t_k,W_{t_k}) \Delta t
\end{equation}
with
\begin{multline*}
c^\alpha_{(k)ab} = 
 \frac{1}{2} \big[ \sigma^\gamma{}_b(t_k,X_{t_k}) e^a{}_\beta(t_k,X_{t_k}) \partial_\gamma C^{\alpha\beta}(t_k,X_{t_k})  
\\
+ \sigma^\beta{}_a(t_k,X_{t_k}) e^b{}_\gamma(t_k,X_{t_k}) \partial_\beta C^{\alpha\gamma}(t_k,X_{t_k})  
\\
-  e^a{}_\beta(t_k,X_{t_k}) e^b{}_\gamma(t_k,X_{t_k}) C^{\alpha\delta}(t_k,X_{t_k}) \partial_\delta C^{\beta\gamma}(t_k,X_{t_k})  \big] \p
\end{multline*}
Equations \eqref{eq:hatMuCr} read
\begin{eqnarray*}
\hat \mu_k^\alpha(\Delta t, \Delta W) &=& f^\alpha_{10} + \frac{1}{2} (f^\alpha_{02})_{bb} + (f^\alpha_{11})_a \Delta W^a
- (g_{11})_a  e^\alpha{}_a(\Delta t, \Delta W) \Delta t
\\
\hat C_k^{\alpha\beta}(\Delta t, \Delta W) &=& e^\alpha_{(k)a}(\Delta t, \Delta W) e^\beta_{(k)a}(\Delta t, \Delta W)
\\
\hat r_k(\Delta t, \Delta W) &=& g_{10} + (g_{11})_a \Delta W^a -\frac{1}{2} (g_{11})_a (g_{11})_a \Delta t^2 \p
\end{eqnarray*}
Using the definitions of coefficients or the fact these quantities should coincide with $\mu^\alpha$, $C^{\alpha\beta}$ and $r$ up to $O(\Delta t)$ terms we can rewrite this as
\begin{eqnarray*}
\hat \mu_k^\alpha(\Delta t, \Delta W) &=& \mu^\alpha(t_k,X_{t_k}) 
+ \sigma^\beta{}_a(t_k,X_{t_k}) 
\\ && \qquad\qquad
\big[ \partial_\beta \mu^\alpha(t_k,X_{t_k}) \Delta W^a
- \partial_\beta r(t_k,X_{t_k})  e^\alpha_{(k)a}(\Delta t, \Delta W) \Delta t \big]
\\
\hat C_k^{\alpha\beta}(\Delta t, \Delta W) &=& e^\alpha_{(k)a}(\Delta t, \Delta W) e^\beta_{(k)a}(\Delta t, \Delta W)
\\
\hat r_k(\Delta t, \Delta W) &=& r(t_k,X_{t_k}) +  \sigma^\alpha{}_a(t_k,X_{t_k})  \partial_\alpha r(t_k,X_{t_k}) \Delta W^a
\\ &&\qquad\qquad\qquad
-\frac{1}{2} C^{\alpha\beta}(t_k,X_{t_k}) \big( \partial_\alpha r(t_k,X_{t_k}) \big)\big(\partial_\beta r(t_k,X_{t_k})\big) \Delta t^2 \p
\end{eqnarray*}
Using these expressions for $k-1$ we compute
\begin{eqnarray*}
\Delta \mu_k^\alpha &=& \mu^\alpha(t_k,X_{t_k}) - \hat \mu_{k-1}^\alpha(\Delta t_{k-1}, \Delta W_{k-1})
\\
\Delta C_k^{\alpha\beta} &=& C^{\alpha\beta}(t_k,X_{t_k}) - \hat C_{k-1}^{\alpha\beta}(\Delta t_{k-1}, \Delta W_{k-1})
\\
\Delta r_k &=& r(t_k,X_{t_k}) - \hat r_{k-1}(\Delta t_{k-1}, \Delta W_{k-1})
\end{eqnarray*}
and we get
$$
1 + \frac{\Delta \mathcal{H}_k}{\lambda} = 1 - \frac{\Delta r_k}{\lambda} + \frac{\Delta \mu^\alpha_k}{\lambda} \partial_{X^\alpha_{t_k}} + \frac{1}{2} \frac{\Delta C^{\alpha\beta}_k}{\lambda} \partial_{X^\alpha_{t_k}}\partial_{X^\beta_{t_k}}
$$

Except on the last time $t_p$, $\Delta \mathcal{H}_k$ acts on
\begin{equation}
\label{eq:intUcorrectiveMulti}
 u_k(t_k,X_{t_k}) = \int \ud X_{t_{k+1}} \widehat{U}_{t_k,t_{k+1}}(X_{t_k},X_{t_{k+1}}) \left(1+\frac{\Delta \mathcal{H}_{k+1}}{\lambda}\right) u_{t_{k+1}}(X_{t_{k+1}}) \p
\end{equation}
More generally we will consider the action on this expression of a second order differential operator
$$
\mathcal{A}_k = A_k + A_k^\alpha \partial_{X^\alpha_{t_k}} + \frac{1}{2} A_k^{\alpha\beta} \partial_{X^\alpha_{t_k}}\partial_{X^\beta_{t_k}} \p
$$

As in section \ref{sec:1dintermediate},e consider the change of variable between $X_t$ and $W_t$ defined by
$$
X^\alpha_t = X^\alpha_{t_k} + f_{(k)}^\alpha(t-t_k,W_t-W_{t_k}) \p
$$
On derivatives it induces
\begin{eqnarray}
\label{eq:multiChangeDer}
\partial_a &=& e^{\alpha}_{(k)a} \partial_\alpha
\\
\nonumber
\partial_a \partial_b &=& e^{\alpha}_{(k)a} e^{\beta}_{(k)b} \partial_\alpha \partial_\beta + c^\alpha_{(k)ab} \partial_\alpha
\end{eqnarray}
and their inverse relations
\begin{eqnarray*}
\partial_\alpha &=& e^{a}_{(k)\alpha} \partial_a
\\
\partial_\alpha \partial_\beta &=& e^{a}_{(k)\alpha} e^{b}_{(k)\beta} \partial_a \partial_b - e^{a}_{(k)\alpha} e^{b}_{(k)\beta} e^{c}_{(k)\gamma} c^\gamma_{(k)ab} \partial_c \p
\end{eqnarray*}
In the new variables, the differential operator $\mathcal{A}_k$ is
\begin{multline*}
\mathcal{A}_k = A_k + A_k^\alpha  e^{a}_{(k)\alpha}(0,0) \partial_{W_{t_k}^a}
- \frac{1}{2} A_k^{\alpha\beta} e^{a}_{(k)\alpha}(0,0) e^{b}_{(k)\beta}(0,0) e^{c}_{(k)\gamma}(0,0) c^\gamma_{(k)ab} \partial_{W^c_{t_k}}
\\
+ \frac{1}{2} A_k^{\alpha\beta} e^{a}_{(k)\alpha}(0,0) e^{b}_{(k)\beta}(0,0) \partial_{W^a_{t_k}}\partial_{W^b_{t_k}} \p
\end{multline*}

This operator acts on expression \eqref{eq:intUcorrectiveMulti}. In the new variables, the evolution operator becomes
\begin{equation}
\label{eq:hatUmulti}
\widehat{U}^{(W)}_{t_k,t_{k+1}}(W_{t_k},W_{t_{k+1}}) = e^{\textstyle -g_{(k)}(t_{k+1}-t_k,W_{t_{k+1}}-W_{t_k})} \varphi(t_{k+1}-t_k, W_{t_{k+1}}-W_{t_k})
\end{equation}
where $\varphi$ is the $d$-dimensional Gaussian kernel
$$
\varphi(\Delta t, \Delta W) = \frac{1}{(2\pi \Delta t)^{d/2}} e^{\textstyle - \frac{1}{2} \frac{\Delta W^a \Delta W^a}{\Delta t}} \p
$$

When acting on the term in $\frac{\Delta \mathcal{H}_{k+1}}{\lambda}$, we differentiate $\widehat{U}^{(W)}_{t_k,t_{k+1}}(W_{t_k},W_{t_{k+1}})$ with respect to $W_{t_k}$, which means multiplying by the weights $\widehat{\mathcal{W}}_{(k)a}$ and $\widehat{\mathcal{W}}_{(k)ab}$, according to their definition 
\begin{eqnarray*}
\partial_{W_{t_k}^a} \widehat{U}^{(W)}_{t_k,t_{k+1}}(W_{t_k},W_{t_{k+1}}) &=&  \widehat{\mathcal{W}}_{(k)a}(\Delta t_k, \Delta W_k)
\widehat{U}^{(W)}_{t_k,t_{k+1}}(W_{t_k},W_{t_{k+1}})
\\
\partial_{W_{t_k}^a}\partial_{W_{t_k}^b} \widehat{U}^{(W)}_{t_k,t_{k+1}}(W_{t_k},W_{t_{k+1}}) &=&  \widehat{\mathcal{W}}_{(k)ab}(\Delta t_k, \Delta W_k) 
\widehat{U}^{(W)}_{t_k,t_{k+1}}(W_{t_k},W_{t_{k+1}})\p
\end{eqnarray*}
From expression \eqref{eq:hatUmulti} we have
\begin{eqnarray*}
\widehat{\mathcal{W}}_{(k)a(\Delta t, \Delta W)} &=& \frac{\Delta W^a}{\Delta t} + \partial_a g_{(k)}(\Delta t, \Delta W) 
\\
&=& \frac{\Delta W^a}{\Delta t} + \sigma^\alpha{}_a(t_k,X_{t_k})  \partial_\alpha r(t_k,X_{t_k}) \Delta t
\end{eqnarray*}
and
\begin{eqnarray*}
\widehat{\mathcal{W}}_{(k)ab}(\Delta t, \Delta W) &=&
\left[ \frac{\Delta W^a}{\Delta t} + \partial_a g_{(k)}(\Delta t, \Delta W)  \right]
\left[ \frac{\Delta W^b}{\Delta t} + \partial_b g_{(k)}(\Delta t, \Delta W)  \right]
\\ && \qquad\qquad\qquad\qquad\qquad\qquad
- \frac{\delta_{ab}}{\Delta t} - \partial_a\partial_b g_{(k)}(\Delta t, \Delta W)
\\
&=&
\left[ \frac{\Delta W^a}{\Delta t} + \sigma^\alpha{}_a(t_k,X_{t_k})  \partial_\alpha r(t_k,X_{t_k}) \Delta t  \right]
\\ && \qquad\qquad\quad
\left[ \frac{\Delta W^b}{\Delta t} + \sigma^\beta{}_b(t_k,X_{t_k})  \partial_\beta r(t_k,X_{t_k}) \Delta t  \right]
- \frac{\delta_{ab}}{\Delta t}  \p
\end{eqnarray*}

For the term where $\widehat{U}{t_k,t{k+1}}$ directly acts on $u_{t_{k+1}}$, we use the fact that 
$\widehat{U}^{(W)}_{t_k,t_{k+1}}(W_{t_k},W_{t_{k+1}})$ given in equation \eqref{eq:hatUmulti} depends only on $W_{t_{k+1}} -W_{t_k} $ to transfer the derivatives from the first variable to the second one:
$$
\partial_{W_{t_k}^a}  \widehat{U}^{(W)}_{t_k,t_{k+1}}(W_{t_k},W_{t_{k+1}}) = -
\partial_{W_{t_{k+1}}^a} \! \widehat{U}^{(W)}_{t_k,t_{k+1}}(W_{t_k},W_{t_{k+1}}) \p
$$
Then we integrate by part on $W_{t_{k+1}}$ to transfer the derivative on $u_{t_{k+1}}$ so that
\begin{multline*}
\int \ud W_{t_{k+1}} \partial_{W_{t_k}^a} \widehat{U}^{(W)}_{t_k,t_{k+1}}(W_{t_k},W_{t_{k+1}}) u_{t_{k+1}}(W_{t_{k+1}})
\\
= - \int \ud W_{t_{k+1}} \Big( \partial_{W_{t_{k+1}}^a}\! \widehat{U}^{(W)}_{t_k,t_{k+1}}(W_{t_k},W_{t_{k+1}}) \Big) u_{t_{k+1}}(W_{t_{k+1}})
\\
= \int \ud W_{t_{k+1}} \widehat{U}^{(W)}_{t_k,t_{k+1}}(W_{t_k},W_{t_{k+1}}) \partial_{W_{t_{k+1}}^a} \!u_{t_{k+1}}(W_{t_{k+1}})
\end{multline*}
and similarly for the second derivative. We then go back to the original variable $X$ at date $t_{k+1}$ using equations \eqref{eq:multiChangeDer}.
 
Assembling all terms, we finally get
$$
\mathcal{A}_k u_{t_k}(X_{t_k}) = \int \ud X_{t_{k+1}} \widehat{U}_{t_k,t_{k+1}}(X_{t_k},X_{t_{k+1}})  \mathcal{A}_{k+1} u_{t_{k+1}}(X_{t_{k+1}})
$$
where we define
$$
\mathcal{A}_{k+1} = A_{k+1} + A_{k+1}^\alpha \partial_{X^\alpha_{t_{k+1}}} + \frac{1}{2} A_{k+1}^{\alpha\beta} \partial_{X^\alpha_{t_{k+1}}}\partial_{X^\beta_{t_{k+1}}}
$$
with
\begin{eqnarray*}
A_{k+1} &=& A_k - d_k(\Delta t_k,\Delta W_k) \frac{\Delta r_{k+1}}{\lambda}
\\
A_{k+1}^\alpha &=&  [\delta^\alpha_\gamma+b_{(k)\gamma}^\alpha] A_k^\gamma
-  e^b{}_{\beta}(t_k,X_{t_k})  e^c{}_{\gamma}(t_k,X_{t_k}) c_{(k)bc}^\delta b_{(k)\delta}^\alpha  A_k^{\gamma\delta}
\\&& \qquad\qquad\qquad\qquad\qquad\qquad\qquad\qquad
+ d_k(\Delta t_k,\Delta W_k) \frac{\Delta \mu^{\alpha}_{k+1}}{\lambda}
\\
A_{k+1}^{\alpha\beta} &=& [\delta^\alpha_\gamma+b_{(k)\gamma}^\alpha][\delta^\beta_\delta+b_{(k)\delta}^\beta] A_{k+1}^{\gamma\delta}
+ d_k(\Delta t_k,\Delta W_k) \frac{\Delta C^{\alpha\beta}_{k+1}}{\lambda} \p
\end{eqnarray*}
We define $b_{(k)\gamma}^\alpha$ by
$$
e^a_{(k)\gamma}(0,0) e^\alpha_{(k)a}(\Delta_k,\Delta W_k)
=
\delta^\alpha_\gamma + b_{(k)\gamma}^\alpha
$$
which in our case, using \eqref{eq:edtdw}
and
$$
e^a_{(k)\gamma}(0,0) = e^a{}_{\gamma}(t_k,X_{t_k}) =\big[\sigma(t_k,W_{t_k})^{-1}\big]^a{}_\gamma \rlap{\ ,}
$$
gives
$$
b_{(k)\gamma}^\alpha = e^a{}_{\gamma}(t_k,X_{t_k}) c^\alpha_{(k)ab} \Delta W_k^b + \partial_\gamma \mu(t_k,X_{t_k}) \Delta t_k \p
$$
We also define the effect of operator $\mathcal{A}_p$ acting by weights multiplication as
\begin{multline*}
d_k(\Delta t,\Delta W) = A_k + A_k^\alpha e^a{}_\alpha(t_k,X_{t_k})  \widehat{\mathcal{W}}_{(k)a}(\Delta t, \Delta W)
\\
+\frac{1}{2} A_k^{\alpha\beta}  e^a{}_{\alpha}(t_k,X_{t_k})  e^b{}_{\beta}(t_k,X_{t_k}) \qquad
\\
\Big[ \widehat{\mathcal{W}}_{(k)ab}(\Delta t, \Delta W)
- e^c{}_{\gamma}(t_k,X_{t_k}) c^\gamma_{(k)ab}  \widehat{\mathcal{W}}_{(k)c}(\Delta t, \Delta W)
\Big] \ .
\end{multline*}

Finally, on the last date, we keep the variance finite by antithetic sampling as in section \ref{sec:payoff}.
We compute
\begin{eqnarray*}
X_T^{\alpha(+)} &=& X^\alpha_{t_p} + f^\alpha_{(p)}(\Delta t_p, \Delta W_p)
\\
X_T^{\alpha(0)} &=& \hat{\mathbb{E}}_{t_p}\big[ X^\alpha_T \big] = X^\alpha_{t_p} + \mu^\alpha(t_p,S_{t_p}) \Delta t_p
\\
X_T^{\alpha(-)} &=& X^\alpha_{t_p} + f^\alpha_{(p)}(\Delta t_p, -\Delta W_p)
\end{eqnarray*}
and the corresponding discount factors on the last time step
\begin{eqnarray*}
D_{p,T}^{(+)} &=& e^{\textstyle - g_{(p)}(\Delta t_p, \Delta W_p)}
\\
D_{p,T}^{(0)} &=& e^{\textstyle - g_{(p)}(\Delta t_p, 0)}
\\
D_{p,T}^{(-)} &=& e^{\textstyle - g_{(p)}(\Delta t_p, -\Delta W_p)} \p
\end{eqnarray*}
Then we can get the contribution from the path to the Monte Carlo estimate as
\begin{multline*}
P_T = D_p \bigg( \frac{1}{2} d_p(\Delta t_p, \Delta W_p) D_{p,T}^{(+)} h\big(S_T^{(+)}\big)
+ \frac{1}{2} d_p(\Delta t_p, -\Delta W_p) D_{p,T}^{(-)} h\big(S_T^{(-)}\big)
\\
- d_p^{(0)} D_{p,T}^{(0)} h\big(S_T^{(0)}\big)
\bigg)
\end{multline*}
with
\begin{eqnarray*}
d_p^{(0)} &=& \frac{1}{2} \Big[ d_p(\Delta t_p, \Delta W_p) + d_p(\Delta t_p, -\Delta W_p) \Big]
- \widehat{\mathbb{E}}_{t_p}\Big[ d_p(\Delta t_p, \Delta W_p) \Big] 
\\
&=& \frac{1}{2} A_p^{\alpha\beta}  e^a{}_{\alpha}(t_k,X_{t_p})  e^b{}_{\beta}(t_p,X_{t_p}) \bigg( \frac{\Delta W_p^a \Delta W_p^b}{\Delta t_p^2} - \frac{\delta_{ab}}{\Delta t_p} \bigg) \p
\end{eqnarray*}
By construction $d_p^{(0)}$ has a null expected value. Thus $P_T$ is the average of two antithetic contributions to the option price, minus a term of null expected value.

Finally, the average of $P_T$ over all paths gives the Monte Carlo estimate of $u_{t_0}(X_0)$. It converges to it when the number of paths goes to infinity without any bias, as explained in section \ref{sec:poisson}.

\subsection{Monte Carlo scheme summary}

In a Monte Carlo simulation, on each path we start by initializing operator $\mathcal{A}_0$ by the identity
$
\mathcal{A}_0 = 1
$
and the disccount factor $D_0$ by 1
$
D_0 = 1
$. We start from $X_{t_0} = X_0$.
We draw Poisson times $t_k$. When going from date $t_k$ to date $t_{k+1}$ on a Monte Carlo path, we do the following:
\begin{enumerate}
\item Compute all coefficients $f_{ij}^\alpha$ and $g_{ij}$ in order to get functions $f_{(k)}^\alpha$ and $g_{ij}$.
\item Draw $d$ independent Gaussian variables $\Delta W_k^a$ with variance $\Delta t_k$ and get the next state $X_{t_{k+1}}$ as
$
X^\alpha_{t_{k+1}} = X^\alpha_{t_k} + f^\alpha_{(k)}(\Delta t_k,\Delta W_k)
$.
\item Accumulate the discount factor as
$
D_{k+1} = D_k e^{\textstyle -g_{(k)}(\Delta t_k,\Delta W_k)}
$.
\item Compute $\mathcal{A}_{k+1}$ from  $\mathcal{A}_k$ as explained above.
\end{enumerate}
When we reach the last time $t_p$ before maturity $T$, we perform steps 1 and 2 with $t_{p+1} = T$, then we compute the corrected, discounted payoff $P_T$ as explained above. The final value is obtained by averaging over all Monte Carlo paths.

\section{Final comments}

In this article, we introduce a Monte Carlo scheme which converges to the theoretical value without any bias, while keeping a finite variance. It applies to multidimensional diffusion processes and it can also handle stochastic interest rates. It allows to decrease the average number of time steps needed to reach a given precision, which can save a lot of computation time.

\subsection{Related work}

We leverage some interesting work presented in \cite{henry2015exact}. However our Monte Carlo scheme differs in several ways.

One of the main differences is that in their scheme, paths which take into account corrective terms do not take into account the basic payoff contribution $h(S_T)$. 
In other words, their choice is equivalent to keeping the constant unit term in equation \eqref{eq:jump} only when there is no jump at time $t$. This occurs with probability $1-\lambda_t \delta t$ and this is compensated by a factor $\frac{1}{1-\lambda_t \delta t} \sim e^{\lambda \delta t}$.  Equation \eqref{eq:jump} is thus replaced by
$$
(1 - \delta N_t)e^{\textstyle \lambda_t \delta t} + \delta N_t \frac{\Delta \mathcal{H}_t}{\lambda_t} \p
$$

The probability of not having any Poisson jump over a maturity $T$ is $e^{-\lambda T}$. Thus only this proportion of the Monte Carlo paths contains the basic payoff contribution. This is compensated by a weight $e^{\lambda T}$ coming from the product of $e^{\lambda \delta t}$ on all infinitesimal times.

However this increases the total Monte Carlo noise, especially for large values of the Poisson intensity $\lambda$. In this case $e^{-\lambda T}$ is close to zero and very few paths, if any, include the payoff contribution $h(S_T)$. In addition all paths contributions include a factor $e^{\lambda T}$ which can become very large.

In our scheme, the payoff contribution is kept for all paths, whether they include corrective terms or not. In addition, there is no such factor $e^{\lambda T}$. This makes the scheme usable for any value of $\lambda$, even when it becomes large.

A second difference is that our simulation scheme can handle simultaneously non-zero drift and nonconstant volatilities, in any dimension.

We also show how to take into account stochastic interest rates.

\subsection{Possible improvements}

The Monte Carlo scheme presented here can be enhanced in several ways. In particular, one can make different choices for the precise form of functions $f^\alpha_{(k)}$ and $g_{(k)}$. Depending on the specific choice, this can allow to have a simulated path closer to the original process and thus the corrective terms would be smaller. A a simple case, one can factor in some time dependence in parameters.

In addition, the Poisson intensity $\lambda$ can depend on time $t$ and on the stochastic variables $X_t^\alpha$. One could increase it in the regions where the corrective terms are higher and decrease it when they are smaller.

\subsection*{Acknowledgment}
We thank Calypso Herrera, Martial Millet and Arnaud Rivoira for useful comments.

\newpage

\bibliographystyle{apalike}
\bibliography{ExactMC}

\end{document}